\documentclass[useAMS,usenatbib]{mn2e}
\usepackage{graphicx}
\usepackage{times}
\usepackage{subfigure}
\usepackage{epsfig}
\usepackage{amssymb}
\usepackage{amsmath}

\title[Distribution Of Clumps In Broad Line Region] {Kinematics and Structure of Clumps in Broad-line Regions in Active Galactic Nuclei}
\author[Mohammad Ghayuri]{Mohammad Ghayuri\thanks{E-mail:
mohammadghayuri@gmail.com} \\
\\Independent Scholar, Mashhad, Iran\\}

\begin{document}

\date{}

\pagerange{\pageref{firstpage}--\pageref{lastpage}} \pubyear{2016}

\maketitle

\label{firstpage}

\begin{abstract}
\\
We use the Jeans equations for an ensemble of collisionless particles to describe the distribution of broad-line region (BLR) cloud in three classes: (A) non disc (B) disc-wind (C) pure disc structure. We propose that clumpy structures in the brightest quasars belong to class A, fainter quasars and brighter Seyferts belong to class B, and dimmer Seyfert galaxies and all low-luminosity AGNs (LLAGNs) belong to class C. We derive the virial factor, $f$, for disc-like structures and find a negative correlation between the inclination angle, $\theta_{0}$, and $f$.  We find similar behaviour for $f$ as a function of the FWHM and $\sigma_{z}$, the $z$ component of velocity dispersion. For different values of $\theta_{0}$ we find that $ 1.0 \lesssim f \lesssim 9.0 $ in type1 AGNs and $ 0.5 \lesssim f \lesssim 1.0 $ in type2 AGNs. Moreover we have $ 0.5 \lesssim f \lesssim 6.5 $ for different values of $\textsc{FWHM}$ and $ 1.4 \lesssim f \lesssim 1.8 $ for different values of $ \sigma_{z} $. We also find that $ f $ is relatively insensitive to the variations of bolometric luminosity and column density of each cloud and the range of variation of $ f $ is in order of 0.01. Considering wide range of $ f $ we see the use of average virial factor $ \langle f \rangle $ is not very safe. Therefore We propose AGN community to divide a sample into a few subsamples based on the value of $\theta_{0}$ and $\textsc{FWHM}$ of members and calculate $ \langle f \rangle $ for each group separately to reduce uncertainty in black hole mass estimation.

\end{abstract}

\begin{keywords}
galaxies: active - galaxies: nuclei - galaxies: Seyfert - galaxies: kinematics and dynamics - black hole physics
\end{keywords}

\section{Introduction}

Now it is widely accepted that an active galactic nucleus (AGN) is a supermassive black hole surrounded by an accretion disc. Above the accretion disc, there is dense, rapidly-moving gas making up the so-called broad line region (BLR) emitting broad emission lines by reprocessing the continuum radiation from the inner accretion disc (see \citealt{Gaskell09}). The BLR is believed to consist of dense clumps of hot gas ($n_H > 10^9$ cm$^{-3}$; $T \sim 10^4$K) in a much hotter, rarefied medium. The motions of the line-emitting clouds is the main cause of the broadening of the line profiles.

The profiles of the broad emission lines is a major source of information about the geometry and kinematics of the BLR. Profiles can be broadly categorized into two shape: single-peaked and double-peaked. It is believed that double-peaked profiles are emitted from disc-like clumpy structure (e.g., \citealp{Chen89b,Chen89a,Eracleous94,Eracleous03,Strateva03}). Although obvious double-peaked profiles are seen in only small fraction of AGNs, this does not mean that such discs are absent in other AGNs. Based on some studies, under the specific conditions, disc-like clumpy structure can even produce single-peaked broad emission lines (e.g., \citealp{Chen89a,Dumont90a,Kollatschny02,Kollatschny03}). Spectropolarimetric observations (\citealt{Smith05}) imply the presence of clumpy discs in BLR.  On the other hand, other authors have suggested a two-component model in which they consider a spherical distribution of clouds surrounding the central black hole in addition to their distribution in the midplane (e.g., \citealp{Popovic04,Bon09}). In this model, while the disc is responsible for the production of the broad wings, the spherical distribution is responsible for the narrow cores.  The integrated emission line profile is a combination of wings and cores.

Using the width of the broad Balmer emission lines, $\textsc{FWHM}$, and an effective BLR radius, $r_{BLR}$, which is obtained by either reverberation mapping (RM) (e.g., \citealp{Blandford82,Gaskell86,Peterson93,Peterson04}) or from the relationship between optical luminosity and $r_{BLR}$ (\citealp{Dibai77,Kaspi00,Bentz06}), masses of black holes are estimated from the virial theorem, $M =  fr_{BLR}\textsc{FWHM}^{2}/G$, where $G$ is the gravitational constant and $f$ is the ``virial factor" depending on the geometry and kinematics of the BLR and the inclination angle, $\theta_{0}$. Unfortunately with current technology it is impossible to directly observe the structure of the BLR. Therefore the true value of $ f $ for each object is unknown and we are required to use average virial factor, $ \langle f \rangle $, to estimate the mass of supermassive black hole. Comparison of virial masses with independent estimates of black hole masses using to $M - \sigma$ relationship (see \citealt{Kormendy13}) have given empirical estimates of the value of $ \langle f \rangle $. However each study has found different value for $ \langle f \rangle $. For example \citet{Onken04} calculate $ \langle f \rangle = 5.5 \pm 1.8 $, \citet{Woo10} calculate $ \langle f \rangle = 5.2 \pm 1.2 $, \citet{Graham11} calculate $ \langle f \rangle = 3.8^{+0.7}_{-0.6} $ and \citet{Grier13} calculate $ \langle f \rangle = 4.31 \pm 1.05 $. This is because each group take a different sample. Different values for $ \langle f \rangle $ prevent us to have a reliable value for black hole mass.

The situation for low-luminosity AGNs (LLAGNs) is somewhat different. The lack of broad optical emission lines in the faintest cases ($ L = 10^{-9}< L <10^{-6} L_{Edd}$) has led to two scenarios about the presence of BLR in LLAGNs. The first, which is somewhat supported by the theoretical models (e.g., \citealp{Nicastro00,Elitzur06}), simply says that the BLR is absent in such faint objects. However, there is clear evidence in favor of the presence of the BLR in some LLAGNs, at least those with $ L > 10^{-5} L_{Edd}$.  This supports a second scenario which says that the BLR exists in LLAGNs but we cannot detect them in the faintest cases because the intensity of their broad emission lines is below the detection threshold set mostly by starlight in the host galaxy. In the Palomar survey, it was found that broad H$\alpha$ emission is present in a remarkably high fraction of LINERs (LLAGNs)(see \citealt{Ho08}). Moreover, double-peaked broad emission lines have been found in some LINERs including NGC~7213 (\citealt{Filippenko84}), NGC~1097 (\citealt{Storchi93}), M81 (\citealt{Bower96}), NGC~4450 (\citealt{Ho00}) and NGC~4203 (\citealt{Shields00}). Other studies have also shown the presence of variable broad emission lines for NGC~1097 (\citealt{Storchi93}), M81 (\citealt{Bower96}) and NGC~3065 (e.g., \citealt{Eracleous01}). Recently, \citet{Balmaverde14} found other LLAGNs with $ L=10^{-5} L_{Edd}$ showing BLRs.

Since the optical spectra of LLAGNs are severely contaminated by the host galaxy, some authors have suggested using the widths of the Paschen lines rather than Balmer lines to determine a $\textsc{FWHM}$ (\citealp{Landt11,Landt13,La Franca15}). Also, in order to estimate the BLR radius, $r_{BLR}$, we can use near-IR continuum luminosity at 1 $\mu m$ (\citealp{Landt11,Landt13}) or X-ray luminosity (\citealp{Greene10}) rather than the optical luminosity.

\citet{Whittle86} used the Boltzmann equation to describe the kinematics of BLR. More recently, \citet{Wang12} showed that the BLR can be considered as a collisionless ensemble of particles. By considering the Newtonian gravity of the black hole and a quadratic drag force, they used the collisionless Boltzmann equation (CBE) to study the dynamics of the clouds for the case where magnetic forces are unimportant. Following this approach, some authors included the effect of magnetic fields on the dynamics of the clouds in the BLR (e.g., \citealt{Khajenabi14}). In this paper, we use the CBE to describe the distribution of the clouds in the BLR. The structure of this paper is as follows: in the section (\ref{s2}) we will establish our basic formalism and apply it in order to classify the clumpy structure of the BLR. In the section (\ref{s3}) we will concentrate on LLAGNs and give more details about the distribution of the clouds in such systems. Moreover, we will derive the virial factor $f$ for them and in the final section, the conclusions are summarized.

\section{Kinematic Equations of Clouds in the BLR}\label{s2}

This paper only considers axisymmetric, steady-state systems. In the subsection (\ref{ss21}) we start by deriving the general form of the Jeans equations in cylindrical coordinates ($R, \phi, z$) for a system of particles subject to velocity-dependent forces as well as position-dependent forces. Then, by assuming axisymmetry ($\partial /\partial \phi =0$) and a steady state ($\partial /\partial t =0$), we simplify the Jeans equations and use them to describe the distribution of the clouds in the BLR.

\subsection{Jeans equations}\label{ss21}
If we define the distribution function $F$ as $F=\Delta n/\Delta x \Delta y \Delta z \Delta^{3} v$, the continuity equation in the phase space (see \citealp[][eq 4.11]{Binney87}) is given by $ \partial F/\partial t + \Sigma_{\alpha = 1}^{6} \partial (F \dot{\omega_{\alpha}})/\partial \omega_{\alpha} = 0 $. This equation can be rewritten as $DF/Dt+F(\partial a_{x}/\partial v_{x}+\partial a_{y}/\partial v_{y}+\partial a_{z}/\partial v_{z})=0 $ in Cartesian position-velocity space ($x,y,z,v_{x},v_{y},v_{z}$) where $DF/Dt$ and $\mathbf{a}$ are respectively the Lagrangian derivative in the phase space and the resultant acceleration vector. On the other hand by the chain rule, the partial derivatives with respect to the Cartesian components of the velocity are related to those respect to the cylindrical components by $ \partial/\partial v_{x}=\cos \phi (\partial/\partial v_{R}) - \sin \phi (\partial/\partial v_{\phi}) $, $ \partial/\partial v_{y}=\sin \phi (\partial/\partial v_{R}) + \cos \phi (\partial/\partial v_{\phi}) $ and $ \partial/\partial v_{z}=\partial/\partial v_{z} $. Also the relationship between the Cartesian and cylindrical components of the acceleration vector is $ a_{x}=a_{R}\cos \phi - a_{\phi} \sin \phi $, $ a_{y}=a_{R}\sin \phi + a_{\phi} \cos \phi $ and $ a_{z}=a_{z} $. Combining all of these equations, we obtain the extended form of the CBE in cylindrical position-velocity space ($R,\phi, z, v_{R}, v_{\phi}, v_{z}$) as
\[ \frac{\partial F}{\partial t}+v_{R} \frac{\partial F}{\partial R}+\frac{v_{\phi}}{R} \frac{\partial F}{\partial \phi}+v_{z} \frac{\partial F}{\partial z}+\left(a_{R}+\frac{v^{2}_{\phi}}{R}\right) \frac{\partial F}{\partial v_{R}}+ \]\[\]
\begin{equation}\label{eq1}
\left(a_{\phi}-\frac{v_{R}v_{\phi}}{R}\right) \frac{\partial F}{\partial v_{\phi}}+a_{z} \frac{\partial F}{\partial v_{z}}+F\left(\frac{\partial a_{R}}{\partial v_{R}}+\frac{\partial a_{\phi}}{\partial v_{\phi}}+\frac{\partial a_{z}}{\partial v_{z}}\right)=0.
\end{equation}
As can be seen, in the absence of velocity-dependent forces, equation (\ref{eq1}) agrees with the standard form (see \citealp[][eq 4.15]{Binney87}). As is shown in the Appendix \ref{a1}, the Jeans equations derived from equation (\ref{eq1}) can be written as
\begin{equation}\label{eq2}
\frac{\partial n}{\partial t} + \frac{1}{R} \frac{\partial}{\partial R} (nR\langle v_{R} \rangle)+\frac{1}{R}\frac{\partial}{\partial \phi} (n\langle v_{\phi} \rangle)+\frac{\partial}{\partial z} (n\langle v_{z} \rangle)=0,
\end{equation}
and
\[\frac{\partial}{\partial t} (n\langle v_{R} \rangle)+\frac{\partial}{\partial R} (n\langle v^{2}_{R} \rangle)+\frac{1}{R}\frac{\partial}{\partial \phi} (n\langle v_{R}v_{\phi} \rangle)+\frac{\partial}{\partial z} (n\langle v_{R}v_{z} \rangle)\]
\begin{equation}\label{eq3}
+n\frac{\langle v^{2}_{R}\rangle -\langle v^{2}_{\phi}\rangle}{R}-n \langle a_{R} \rangle =0,
\end{equation}
and
\[\frac{\partial}{\partial t} (n\langle v_{\phi} \rangle)+\frac{\partial}{\partial R} (n\langle v_{R}v_{\phi} \rangle)+\frac{1}{R} \frac{\partial}{\partial \phi} (n\langle v^{2}_{\phi} \rangle)+\frac{\partial}{\partial z} (n\langle v_{\phi}v_{z} \rangle)\]
\begin{equation}\label{eq4}
+\frac{2n}{R}\langle v_{\phi}v_{R}\rangle -n \langle a_{\phi} \rangle =0,
\end{equation}
and
\[\frac{\partial}{\partial t} (n\langle v_{z} \rangle)+\frac{\partial}{\partial R} (n\langle v_{R}v_{z} \rangle)+\frac{1}{R}\frac{\partial}{\partial \phi} (n\langle v_{\phi}v_{z} \rangle)+\frac{\partial}{\partial z} (n\langle v^{2}_{z} \rangle)\]
\begin{equation}\label{eq5}
+\frac{n\langle v_{R}v_{z}\rangle}{R}-n \langle a_{z} \rangle =0,
\end{equation}
where $n$ is the volume number density in the position-place. Equations (\ref{eq2}) - (\ref{eq5}) are an extended form of the Jeans equations describing a collisionless system of particles undergoing both velocity-dependent and position-dependent forces. Considering gravity as dominant force for a axisymmetric system of particles, equations (\ref{eq2}) - (\ref{eq5}) reduce to the standard form of the Jeans equations (see equations 4.28 and 4.29 of \citealp{Binney87}).

\subsection{Dynamics and geometry of BLR}\label{ss22}

In this subsection, by considering a steady axisymmetric system, we include the Newtonian gravity of the black hole, the isotropic radiation pressure of the central source, and the drag force between the clouds and the ambient medium for the linear regime as the dominant forces. First we discuss about the role of radiation pressure and gravity and after that, through the analysis of the clouds near the midplane, we classify the distribution of the clouds in the BLR.
\subsubsection{Radiation pressure versus gravity}
Assuming that the clouds are optically thick, the radiative force can be expressed as
\begin{equation}\label{eq6}
\mathbf{F}_{rad}=\frac{\sigma}{c}\mathcal{F}\mathbf{e}_{r},
\end{equation}
where $\sigma $ and $ c $ are the cloud's cross-section and the speed of the light respectively, and the isotropic radiation flux, $\mathcal{F}$, is
\begin{equation}\label{eq7}
\mathcal{F}(r)=\frac{L}{4\pi r^{2}}.
\end{equation}
$L$ is the bolometric luminosity of the central source and $ r=\sqrt{R^{2}+z^{2}}$ is the spherical radius. For the clouds near the midplane, $ z \ll R$, the radiative force per unit of mass can be written as
\begin{equation}\label{eq8}
\mathbf{a}_{Rad}=\Omega_{k,mid}^{2}\frac{3l}{2\mu \sigma_{T}N_{cl}}(R\mathbf{e}_{R}+z\mathbf{e}_{z}),
\end{equation}
where $\Omega_{k,mid}=\sqrt{GM/R^{3}}$ is the Keplerian angular velocity in the midplane, $\mu $ is the mean molecular weight, $\sigma_{T}$ is the Thomson cross-section, $l$ is the Eddington ratio, and $ N_{cl}$ is the column density of each cloud. Following previous studies, we consider the clouds, with conserved mass $ m_{cl}$, in pressure equilibrium with the inter-cloud gas (e.g., \citealp{Netzer10,Krause11,Khajenabi15}). Furthermore, the pressure of the ambient medium, and hence the gas density in individual clouds, $n_{gas}$, are assumed to have a power-law dependence on the (spherical) distance from the centre as $n_{gas} \propto r^{-s}$. As a result, since $r \approx R $ for the clouds near the midplane, the column density defined by $ N_{cl}=m_{cl}/R_{cl}^{2}$ finally becomes
\begin{equation}\label{eq9}
N_{cl}=N_{0}\left(\frac{R}{R_{0}}\right)^{-2s/3},
\end{equation}
where $R_{0}$ is one light day and $N_{0}$ is the column density at $R_{0}$.

In addition to the gravitational and radiative forces, the drag force opposing relative movement of clouds and the ambient gas is another force that needs to be taken into consideration. Depending on the size of the clouds, there are two regimes for the drag force.  These include the Epstein and the Stokes regimes (e.g., \citealt{Armitage13}). In the Epstein regime, which dominates for the small clouds, the magnitude of the force is proportional to the relative velocity.  In the Stokes regime the drag affecting the movement of the large clouds increases as the square of the relative velocity. We assume the clouds have spherical shapes, so the drag coefficient in the Stokes regime depends solely on the Reynolds number which is proportional to the relative velocity. \citet{Shadmehri15} demonstrated that the Reynolds number in the inter-cloud gas is lower than unity. This means that: (1) we can consider the inter-cloud gas as having a laminar flow and (2) the drag coefficient in the Stokes regime is proportional to the inverse of the Reynolds number (e.g., \citealt{Armitage13}). This means that the drag coefficient in the Stokes regime is proportional to the inverse of relative velocity. As a result, in both the Epstein and Stokes regimes, the magnitude of the drag force is proportional to the relative velocity and both small and large clouds are affected by a linear drag force as $\mathbf{F_{d}}=f_{l}(\mathbf{v}-\mathbf{w})$, where $\mathbf{w}$ is the velocity of the ambient medium and $f_{l}$ is the drag coefficient. The equations of motion of an individual cloud near the midplane are therefore given by

\[a_{R}=-\Omega_{k,mid}^{2}R\left[1-\left(\frac{R}{R_{c}}\right)^{2s/3}\right]-f_{l}(v_{R}-w_{R}),\]
\[a_{\phi}=-f_{l}(v_{\phi}-w_{\phi}),\]
\begin{equation}\label{eq10}
a_{z}=-\Omega_{k,mid}^{2}z\left[1-\left(\frac{R}{R_{c}}\right)^{2s/3}\right]-f_{l}(v_{z}-w_{z}),
\end{equation}
where $R_{c}$ is the critical radius defined as
\begin{equation}\label{eq11}
R_{c}=R_{0}\left( \frac{2\mu\sigma_{T}N_{0}}{3l}\right)^{3/2s}.
\end{equation}
From equations (\ref{eq8}) and (\ref{eq9}), we see that at $R=R_{c}$, the attractive gravitational and repulsive radiative force cancel each other and the magnitude of the resultant force is equal to zero. For $R > R_{c}$, the radiative force is stronger than gravity and accelerates the clouds away from the central black hole.  For $R < R_{c}$ gravity overcomes the radiative force and pulls the clouds inwards. If we assume $ s=3/2 $ , $ \mu=0.61 $, $ l=0.001 $, $ N_{0}\approx 10^{23} cm^{-2} $ and $ \sigma_{T}=6.7\times 10^{-25} cm^{2} $, the value of $ R_{c} $ almost becomes 27 light days which is in order of BLR radius (e.g., \citealt{Krolik91}). Assuming a steady state and axisymmetry, we substitute equations (\ref{eq10}) into equations (\ref{eq3}) - (\ref{eq5}) to obtain

\begin{figure}
\centering
\includegraphics[width=0.8\columnwidth]{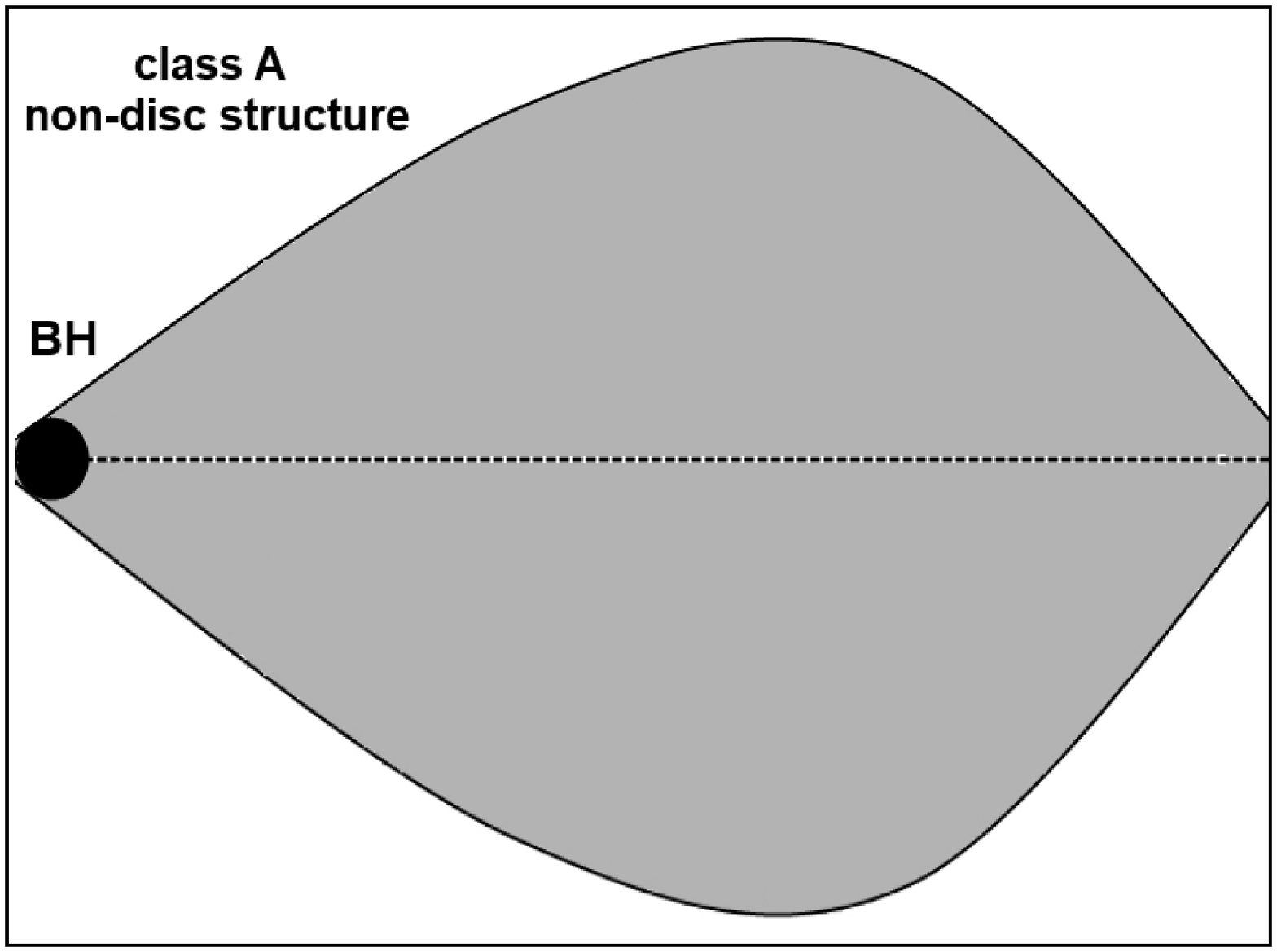}\\
\includegraphics[width=0.8\columnwidth]{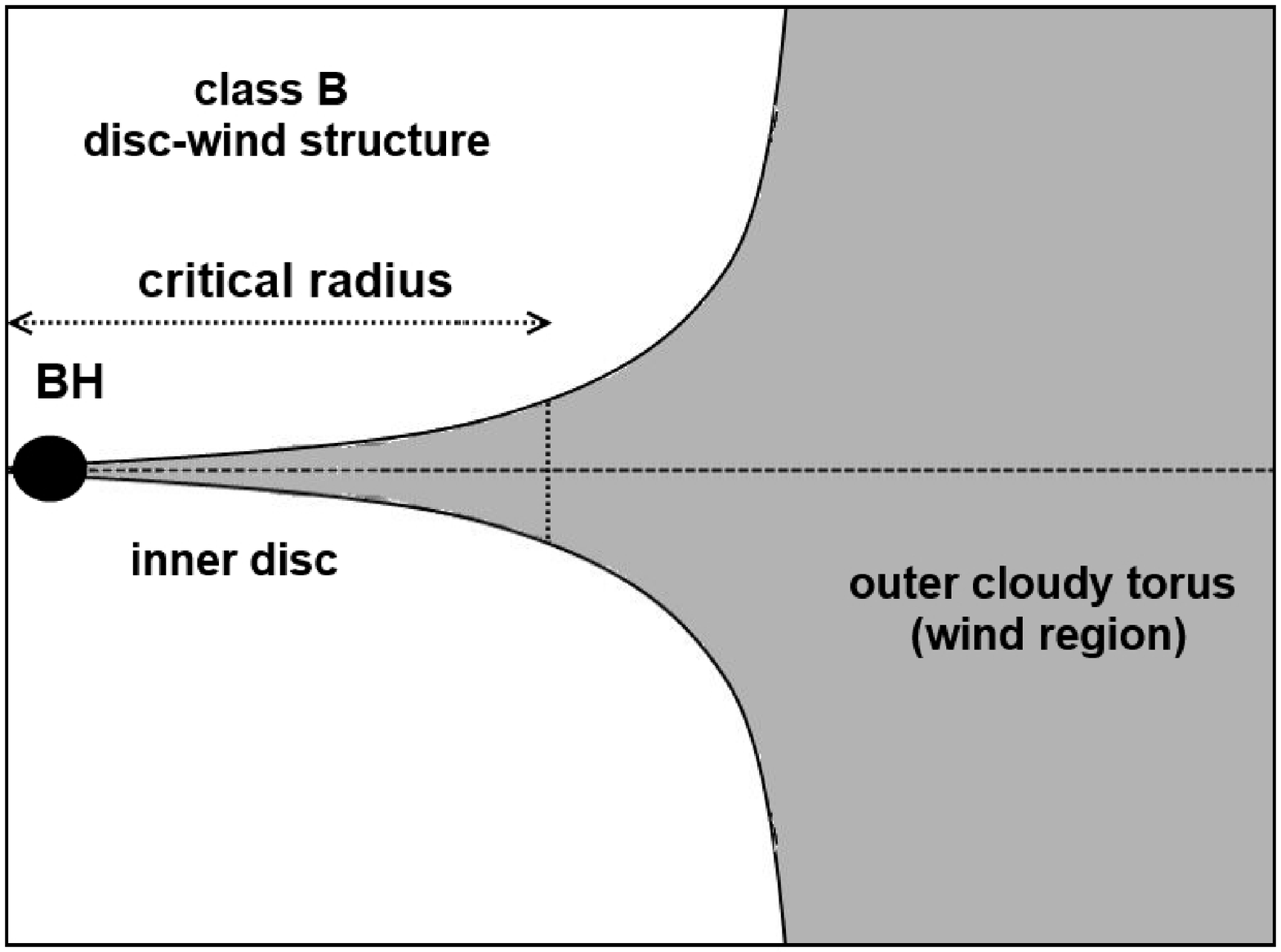}\\
\includegraphics[width=0.8\columnwidth]{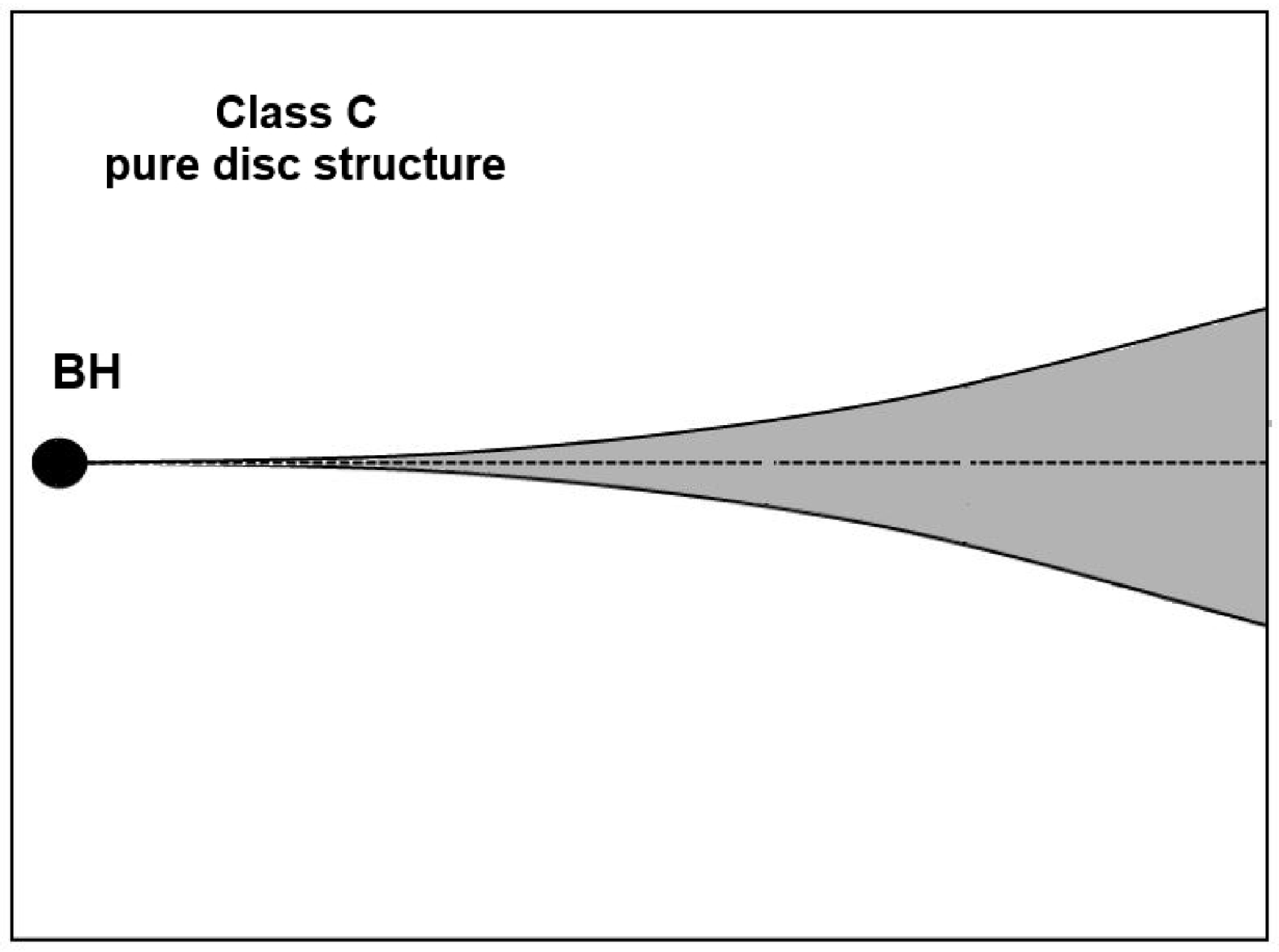}
\caption{Classification of clumpy BLR structure. The black filled circle on the left shows the supermassive black hole. The grey and white areas show the clumpy and non-clumpy regions in the BLR respectively. The first panel shows class A in which the clouds occupy all positions in the BLR. The second panel shows class B. In this class we have a disc for $R < R_{c}$ and a cloudy torus (wind region) for $R > R_{c}$. The third panel shows class C in which the clumpy structure in the BLR is disc-like.}
\label{figure1}
\end{figure}
\begin{figure}
\centering
\includegraphics[width=0.8\columnwidth]{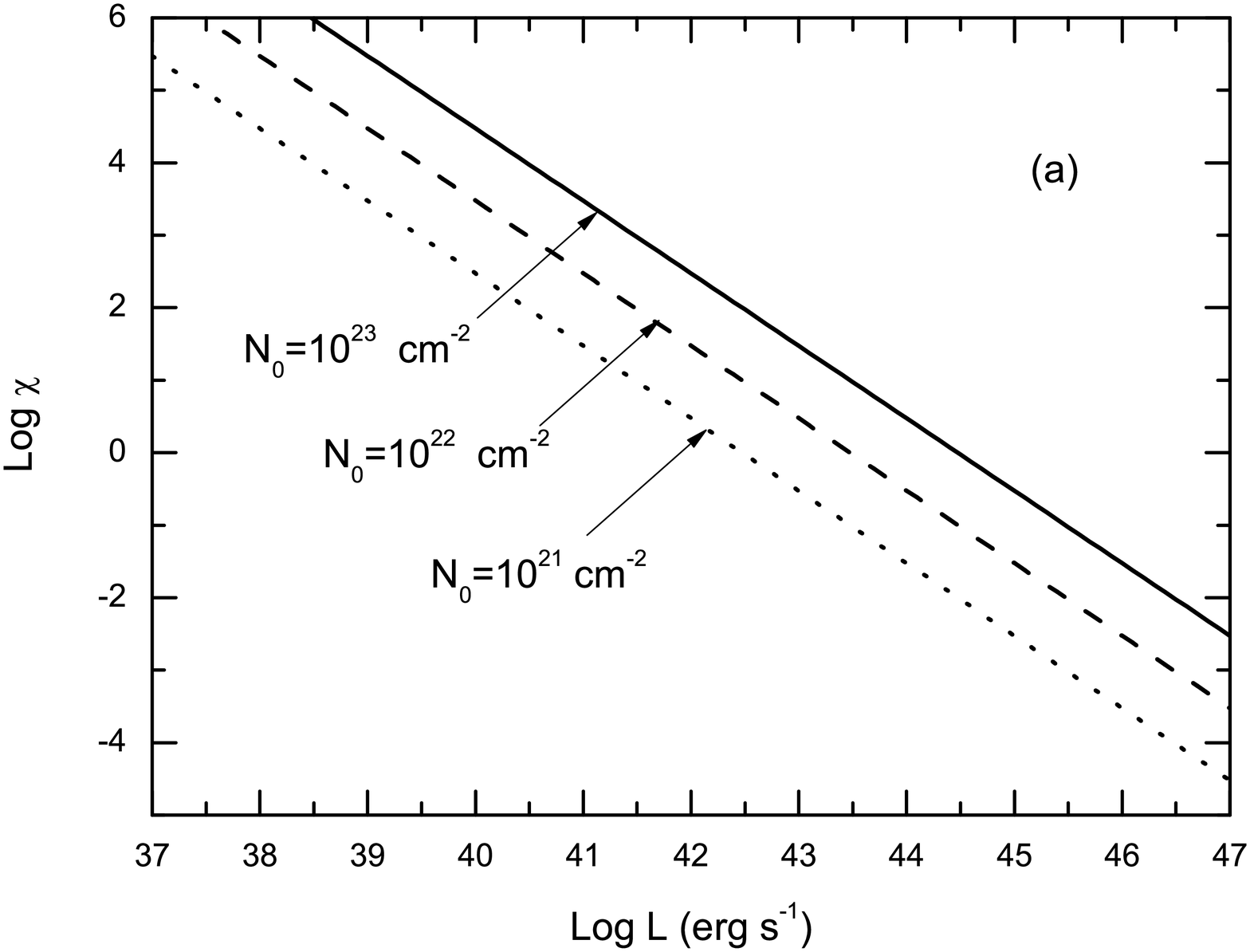}\\
\includegraphics[width=0.8\columnwidth]{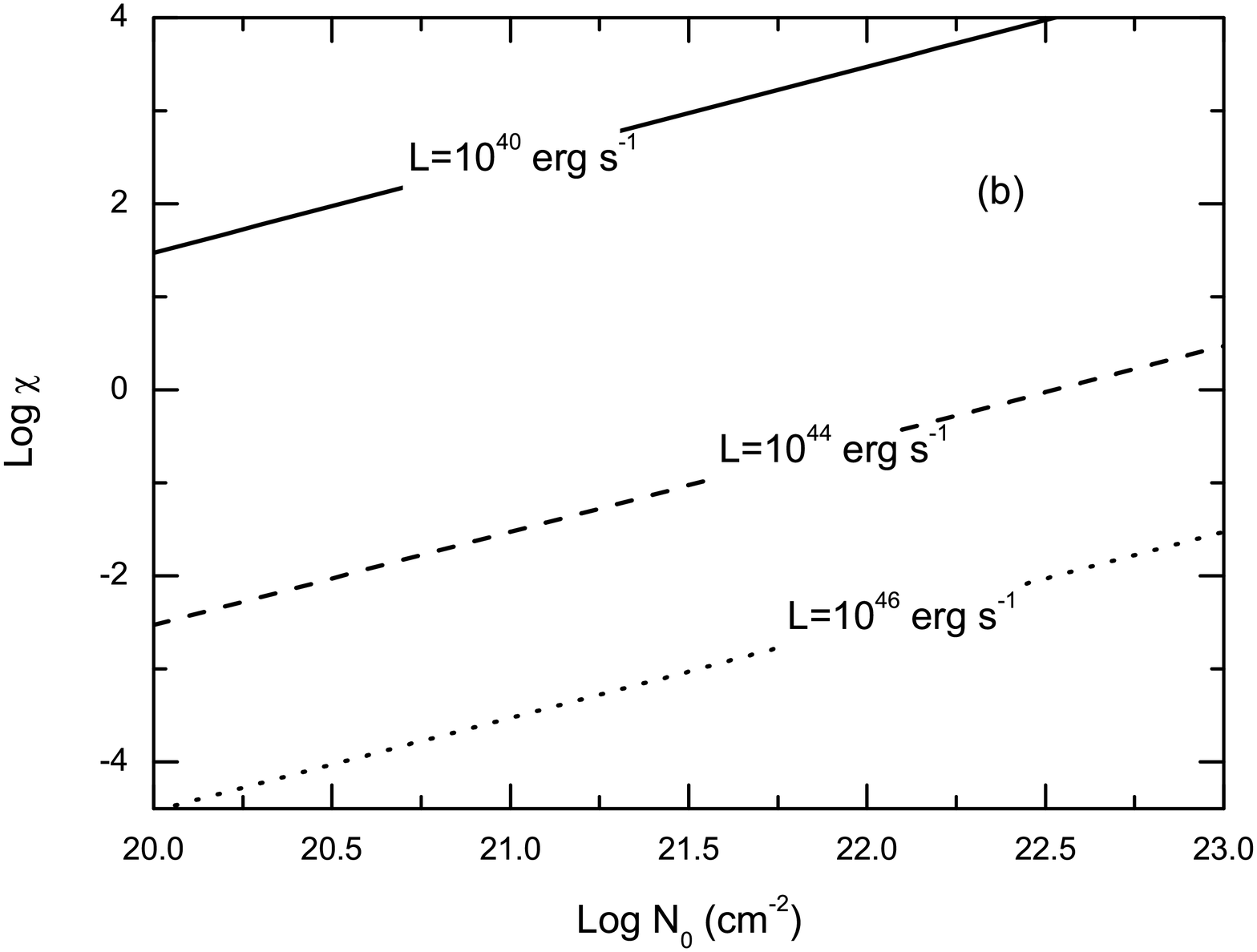}
\caption{(a) $\log \chi$ versus $\log L$, for different values of the column density, $N_{0}$. (b) $\log \chi$ versus $\log N_{0}$ for different values of the bolometric luminosity, $L$. In this Figure, $\log \chi < -2$ and $ -2 < \log \chi < 0$ and $\log \chi > 0$ represents classes A, B, and C respectively. }
\label{figure2}
\end{figure}

\[\frac{\partial}{\partial R} (n\langle v^{2}_{R} \rangle)+\frac{\partial}{\partial z} (n\langle v_{R}v_{z} \rangle)+n\frac{\langle v^{2}_{R}\rangle -\langle v^{2}_{\phi}\rangle }{R}\]
\begin{equation}\label{eq12}
+n\Omega_{k,mid}^{2}R\left[1-\left(\frac{R}{R_{c}}\right)^{2s/3}\right]+nf_{l}(\langle v_{R}\rangle -w_{R}) =0,
\end{equation}
and
\begin{equation}\label{eq13}
\frac{\partial}{\partial R} (n\langle v_{R}v_{\phi} \rangle)+\frac{\partial}{\partial z} (n\langle v_{\phi}v_{z} \rangle)+\frac{2n}{R}\langle v_{\phi}v_{R}\rangle+nf_{l}(\langle v_{\phi}\rangle -w_{\phi}) =0,
\end{equation}
and
\[\frac{\partial}{\partial R} (n\langle v_{R}v_{z} \rangle)+\frac{\partial}{\partial z} (n\langle v^{2}_{z} \rangle)+\frac{n\langle v_{R}v_{z}\rangle}{R}+n\Omega_{k,mid}^{2}z\left[1-\left(\frac{R}{R_{c}}\right)^{2s/3}\right] \]
\begin{equation}\label{eq14}
+nf_{l}(\langle v_{z}\rangle -w_{z}) =0.
\end{equation}
In this paper, we assume that the drag coefficients are sufficiently large that the clouds and the ambient medium are strongly coupled to each other. Thus we can write $\langle v_{R} \rangle=w_{R}$, $\langle v_{\phi} \rangle=w_{\phi}$ and $\langle v_{z} \rangle=w_{z}$.  For simplicity we also assume $\langle v_{z} \rangle =0$, $\langle v_{R}v_{z} \rangle =0 $ and $\langle v_{\phi}v_{z} \rangle =0$. Therefore equations (\ref{eq2}), (\ref{eq12}), (\ref{eq13}) and (\ref{eq14}) are respectively reduced to
\begin{equation}\label{eq15}
\frac{1}{R} \frac{\partial}{\partial R} (nR\langle v_{R} \rangle)=0,
\end{equation}
\begin{equation}\label{eq16}
\frac{\partial}{\partial R} (n\langle v^{2}_{R} \rangle)+n\frac{\langle v^{2}_{R}\rangle -\langle v^{2}_{\phi}\rangle }{R}+n\Omega_{k,mid}^{2}R\left[1-\left(\frac{R}{R_{c}}\right)^{2s/3}\right]=0,
\end{equation}
\begin{equation}\label{eq17}
\frac{\partial}{\partial R} (n\langle v_{R}v_{\phi} \rangle)+\frac{2n}{R}\langle v_{R}v_{\phi}\rangle =0,
\end{equation}
and
\begin{equation}\label{eq18}
\langle v^{2}_{z} \rangle \frac{\partial n}{\partial z} +n\Omega_{k,mid}^{2}z\left[1-\left(\frac{R}{R_{c}}\right)^{2s/3}\right]=0.
\end{equation}
For simplicity $\langle v^{2}_{z} \rangle $ is assumed to be constant in equation (\ref{eq18}). Equation (\ref{eq18}) shows while for $R < R_{c}$ we have $\partial n/\partial z<0 $ and most of the clouds are distributed near the midplane, for $R>R_{c}$ we have $\partial n/\partial z>0 $ and the clouds tend to be located in the higher altitudes. This result is quite in an agreement with what we discussed about $ R_{c} $.
\subsubsection{Classification of clumpy distribution}
\begin{figure}
\centering
\includegraphics[width=0.8\columnwidth]{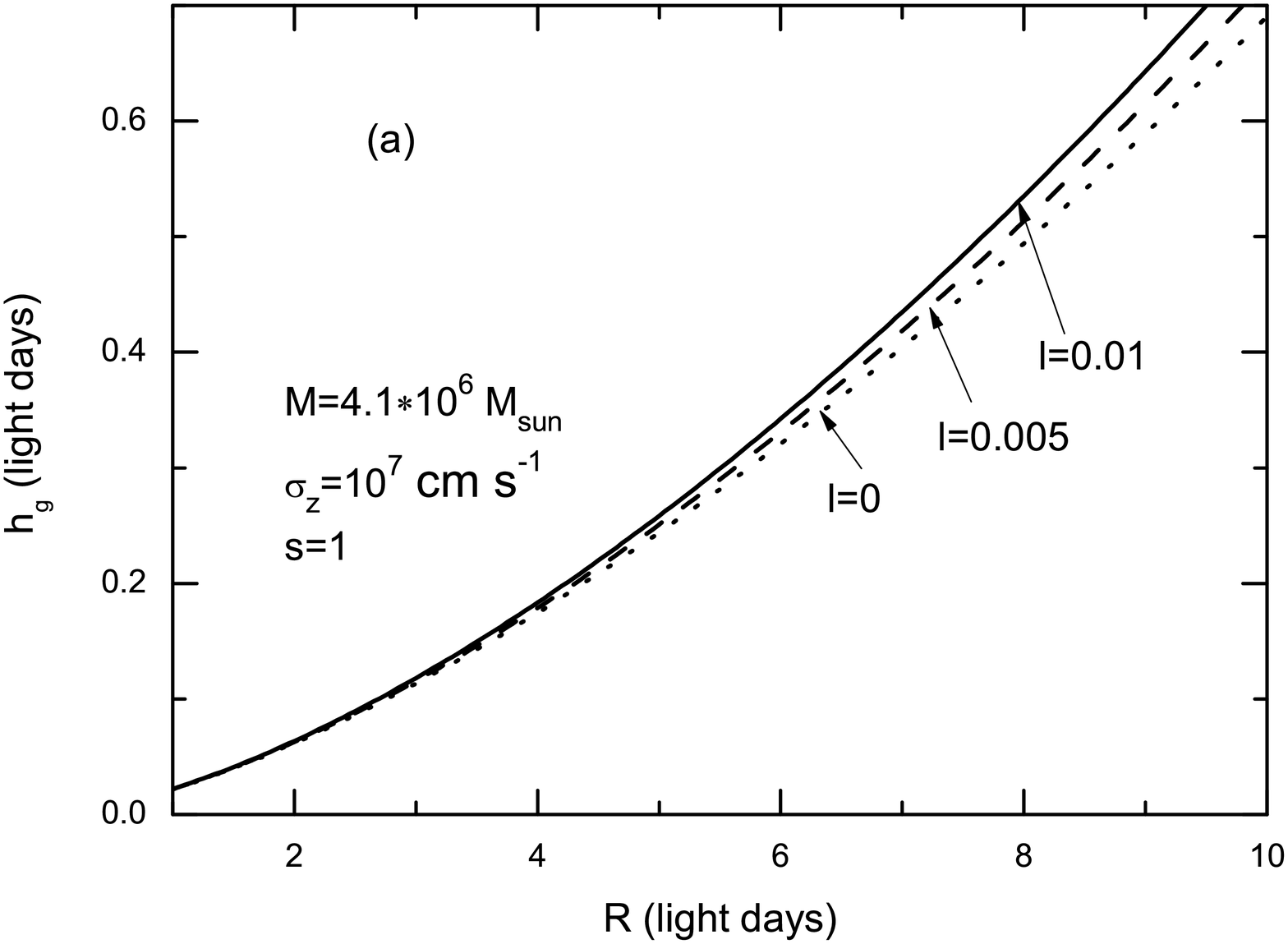}\\
\includegraphics[width=0.8\columnwidth]{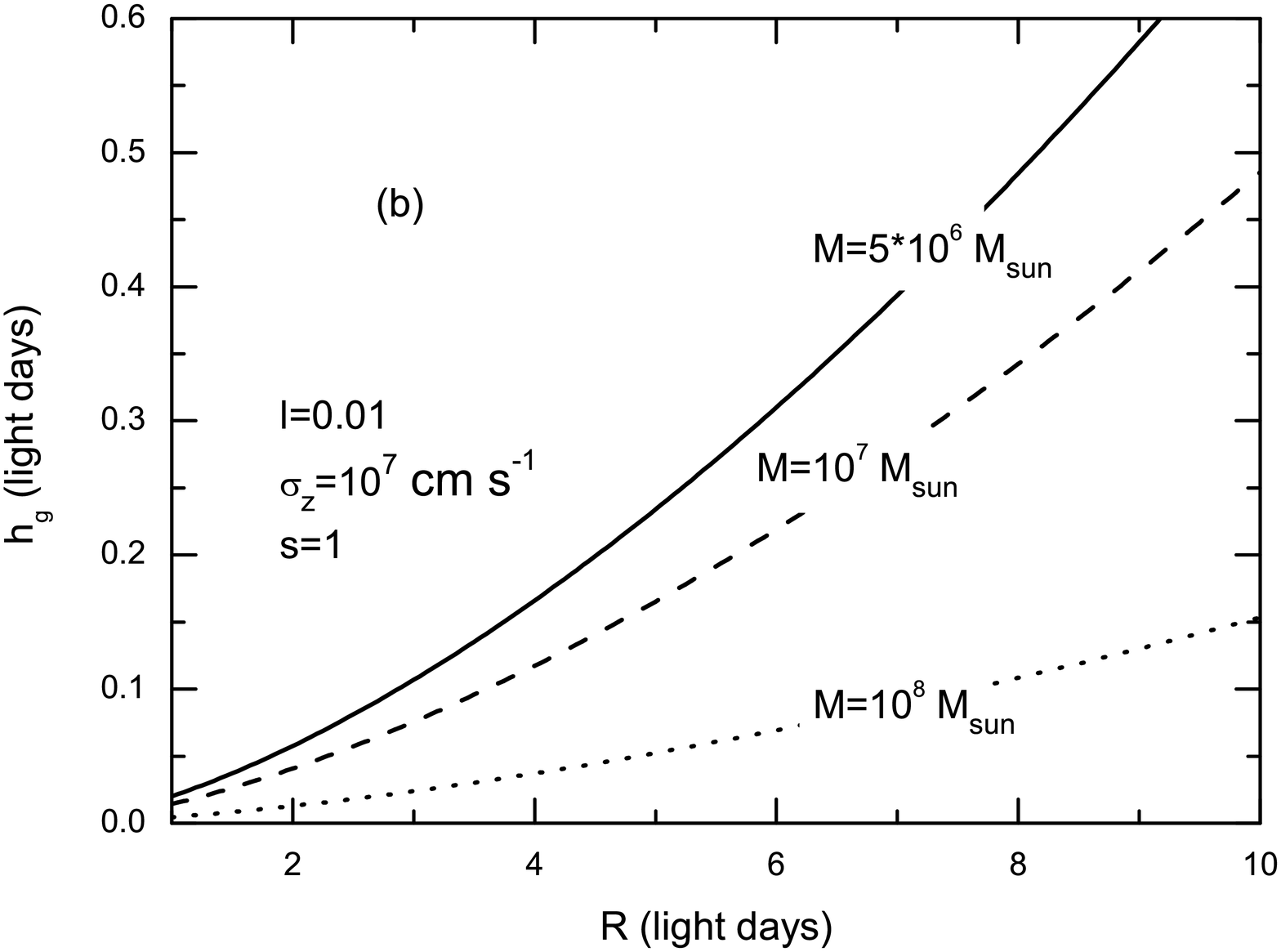}\\
\includegraphics[width=0.8\columnwidth]{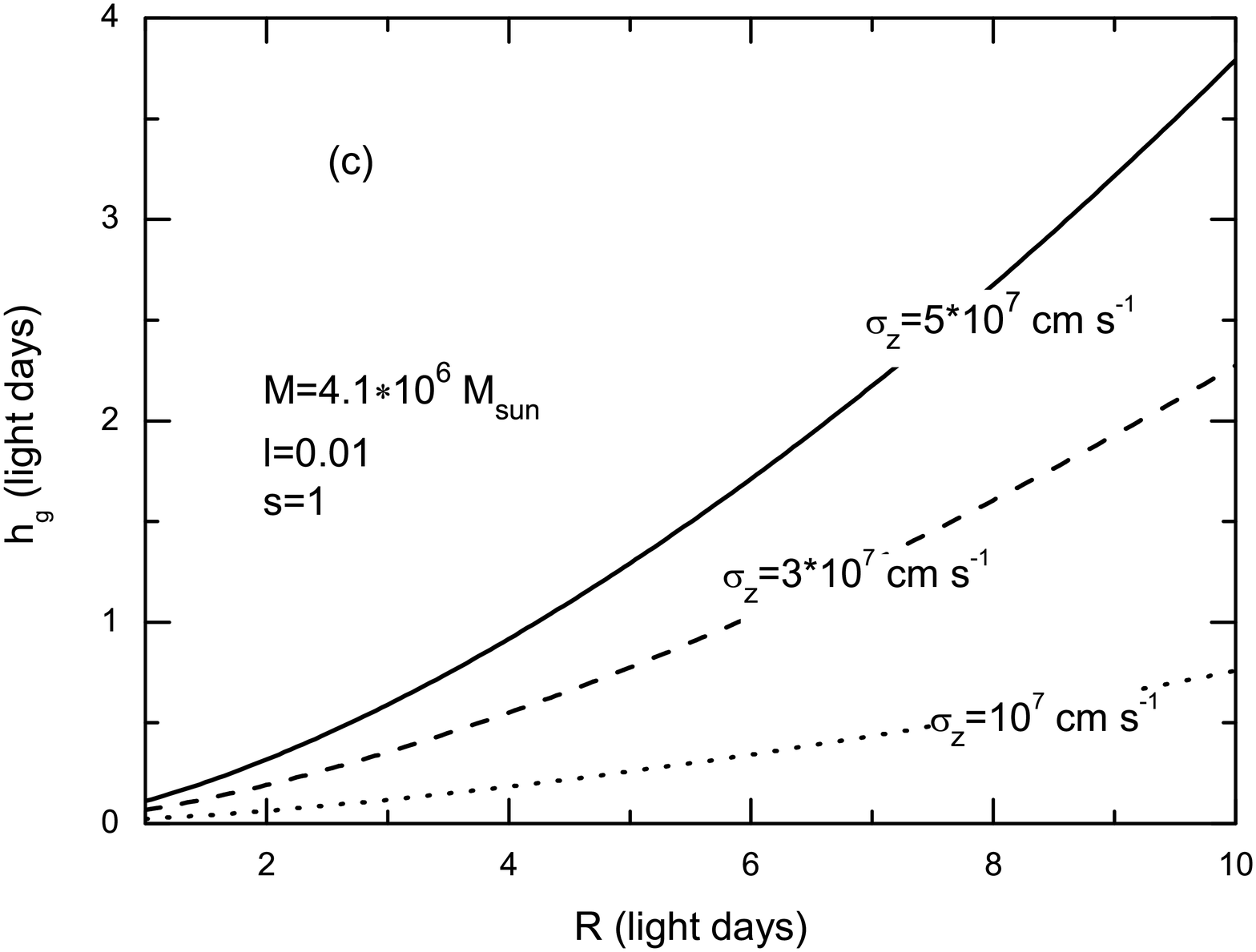}\\
\includegraphics[width=0.8\columnwidth]{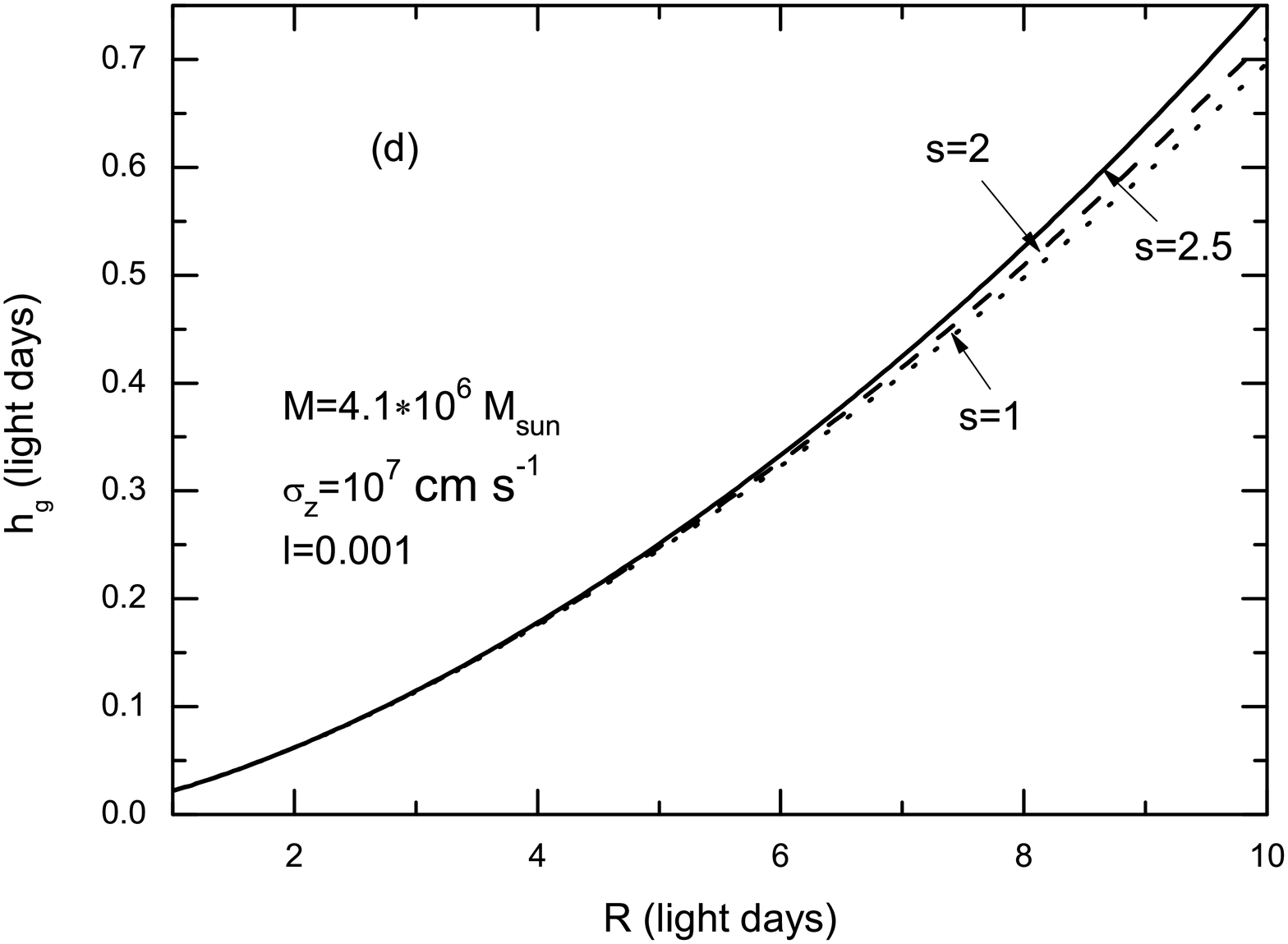}
\caption{The scale height $ h_{g}$ versus the radial distance, $R$. The panels (a), (b), (c) and (d) respectively show the effect of the central luminosity , the black hole mass, z-component of the velocity dispersion and the density index on the geometric shape and thickness of the clumpy disc.}
\label{figure3}
\end{figure}
\begin{figure}
\centering
\includegraphics[width=0.8\columnwidth]{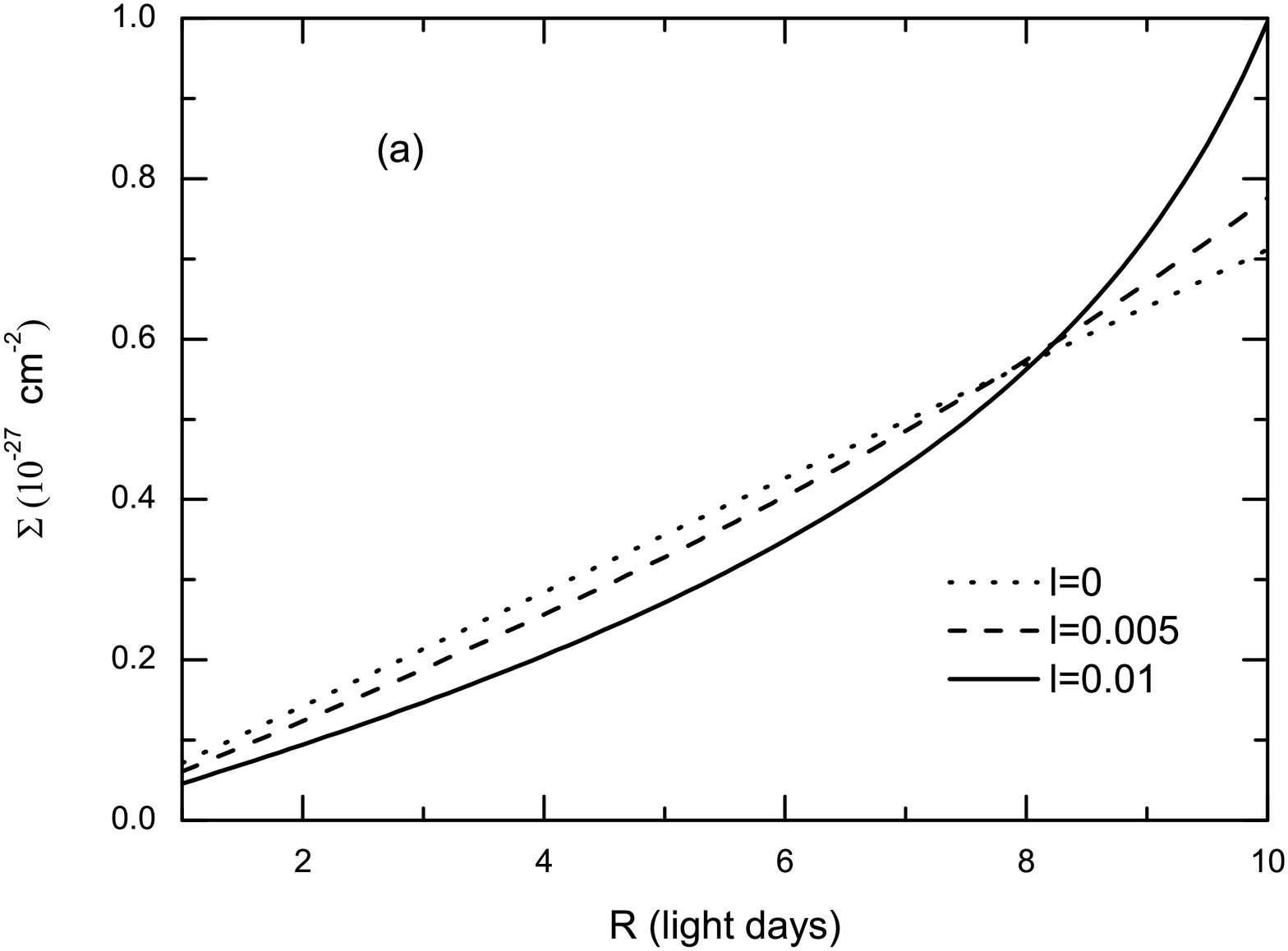}
\includegraphics[width=0.8\columnwidth]{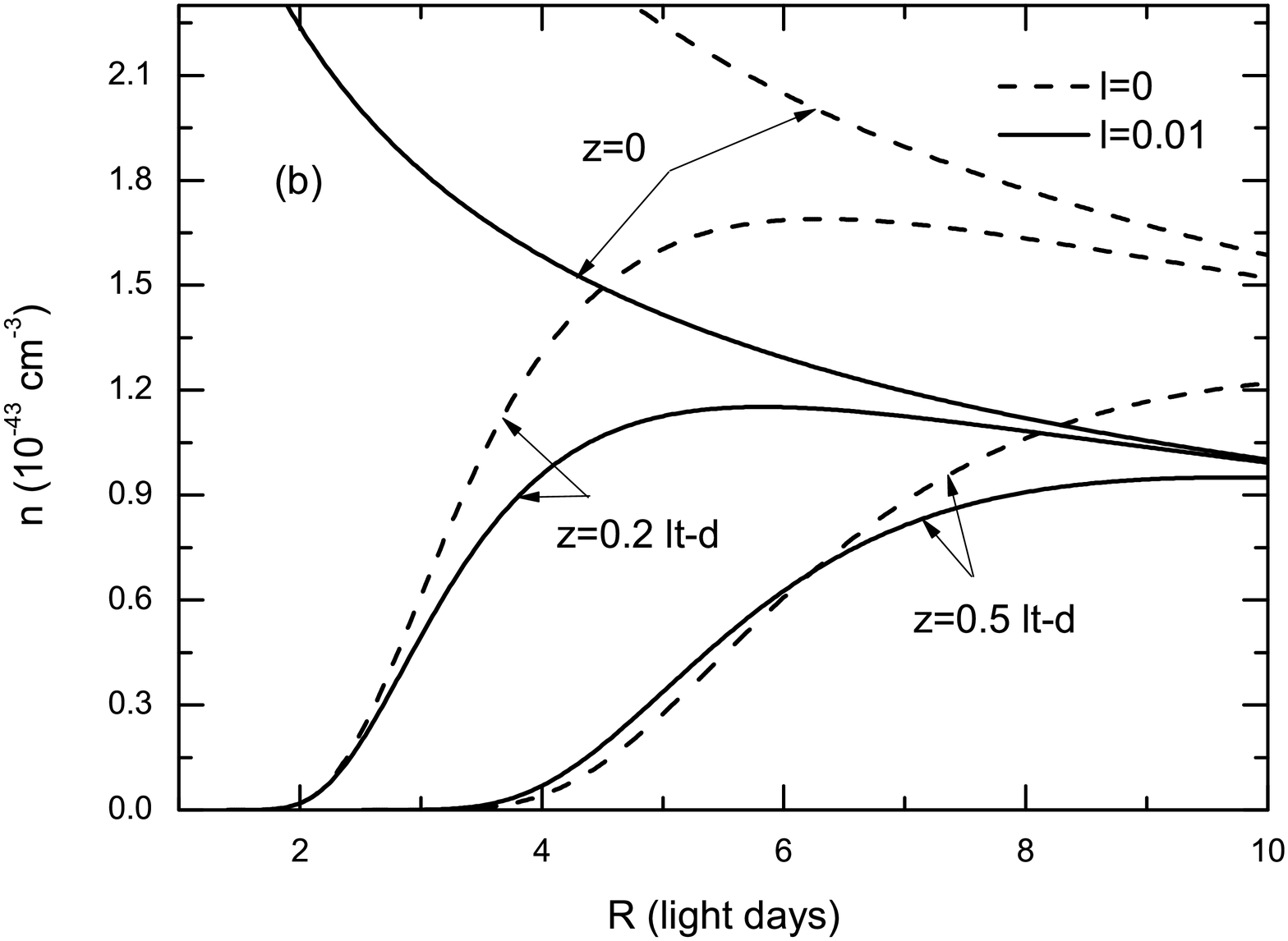}
\caption{The distribution of BLR clouds for different values of the central luminosity. $ M=4.1 \times 10^{6}M_{\odot}$, $\sigma_{z}=10^{7} cm/s$, $ s=1 $ and $ n_{tot}=10^{6}$. (a) The surface number density versus radial distance (b) The volume number density versus radial distance at $ z=0, 0.2 $ and $ 0.5 $ light day (lt-d).}
\label{figure4}
\end{figure}
\begin{figure}
\centering
\includegraphics[width=0.8\columnwidth]{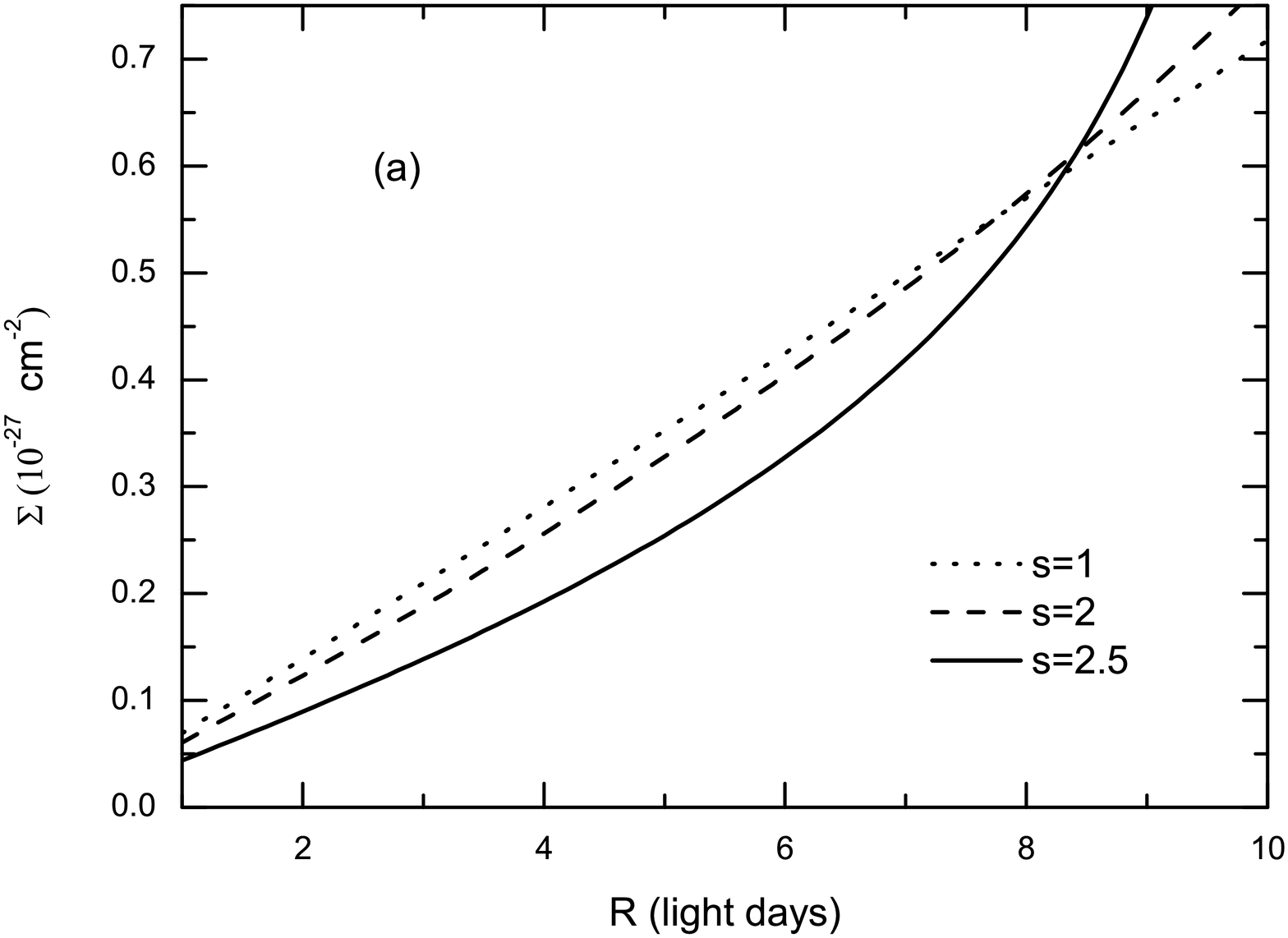}
\includegraphics[width=0.8\columnwidth]{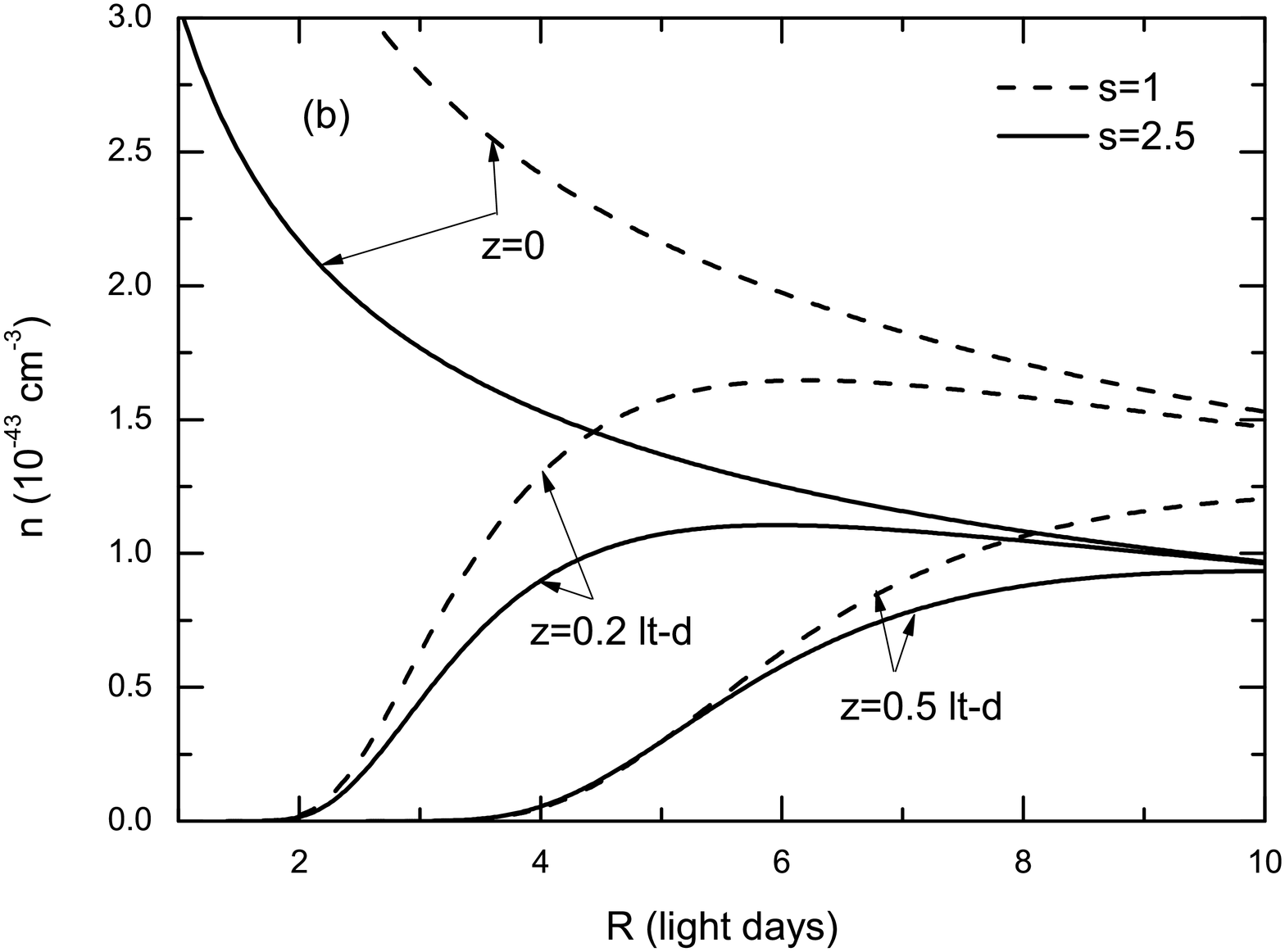}
\caption{The distribution of BLR clouds for different density indices. $M=4.1 \times 10^{6}M_{\odot}$, $\sigma_{z}=10^{7} cm/s$, $ l=10^{-4}$ and $n_{tot}=10^{6}$. (a) The surface number density versus the radial distance, (b) The volume number density versus radial distance at $ z=0, 0.2 $ and $ 0.5 $ light days (lt-d).}
\label{figure5}
\end{figure}
\begin{figure}
\centering
\includegraphics[width=0.8\columnwidth]{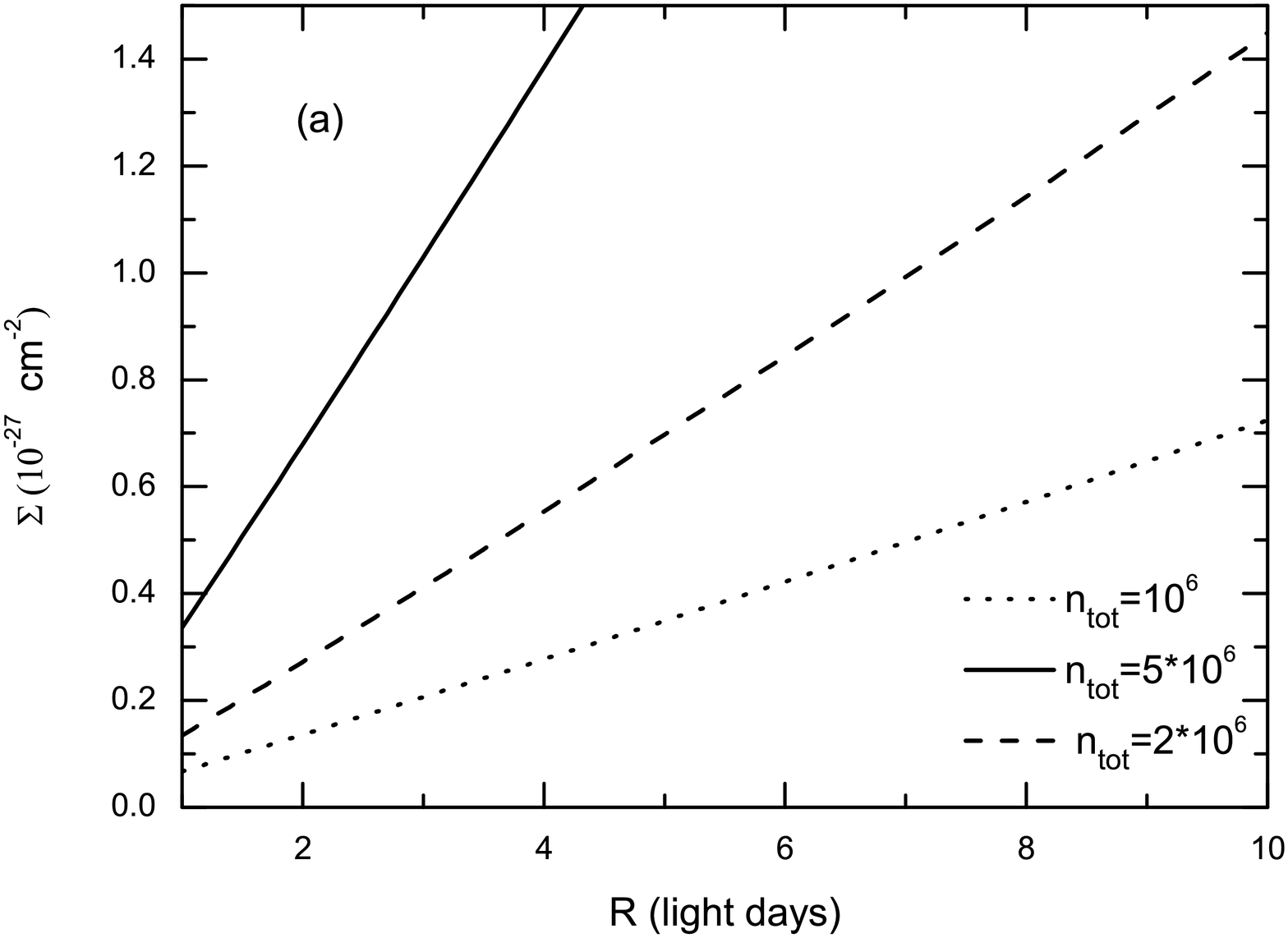}
\includegraphics[width=0.8\columnwidth]{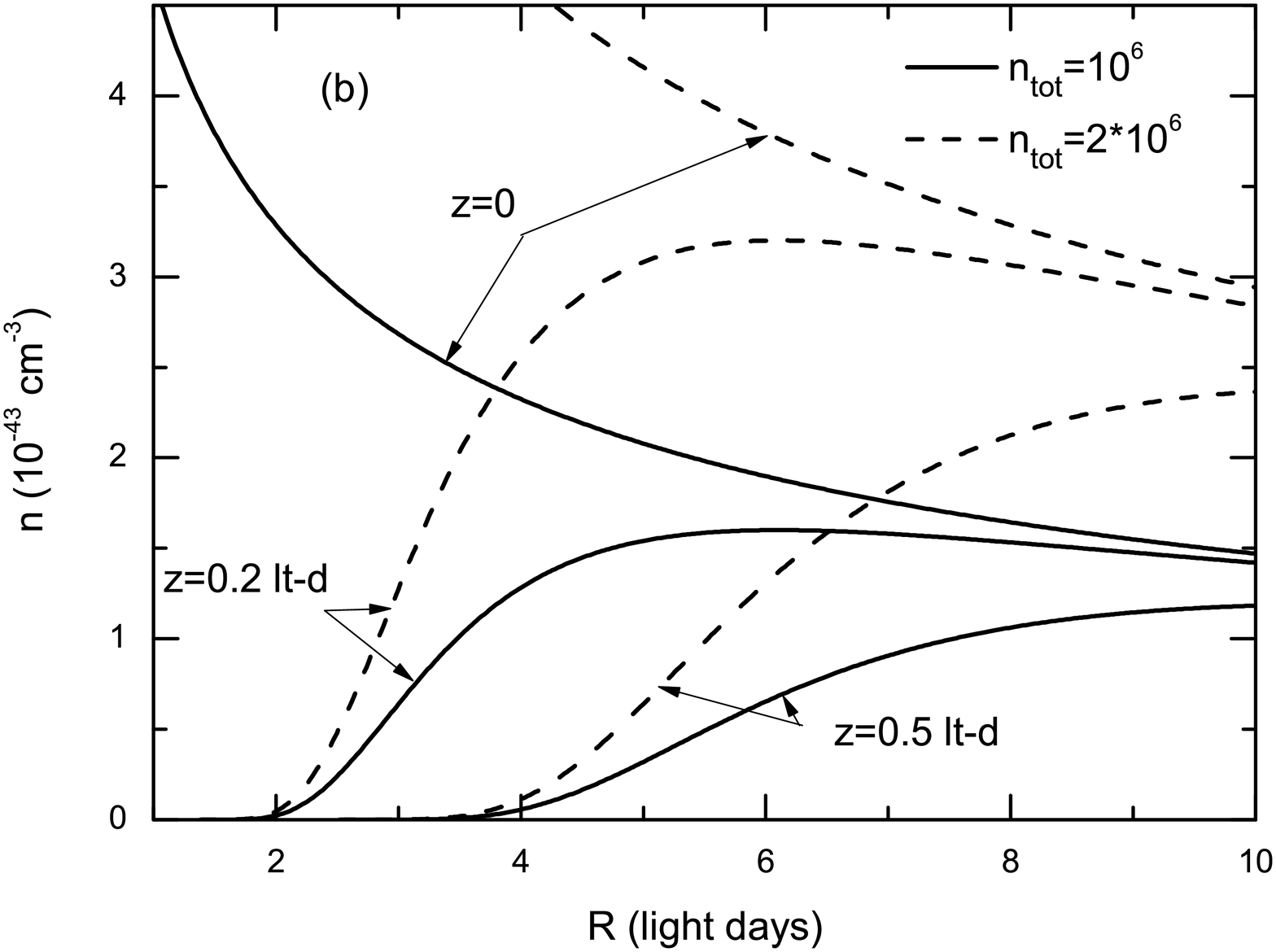}
\caption{The distribution of BLR clouds for different values of the total number of the clouds. $M=4.1 \times 10^{6}M_{\odot}$, $\sigma_{z}=10^{7} cm/s$, $ l=10^{-2}$ and $ s=1 $. (a) Surface number density versus the radial distance (b) Volume number density versus the radial distance at $ z=0, 0.2 $ and $ 0.5 $ light days (lt-d).}
\label{figure6}
\end{figure}

In this part we compare $R_{c}$ with the innermost radius of the BLR $R_{in}$ and the outermost radius $R_{out}$ to show that there are three classes for clumpy distribution in the BLR:

\textbf{Class A}: $R_{c } < R_{in}$ and the clouds fill all positions in the intercloud gas (see the first panel in Figure 1).

\textbf{Class B}: $R_{in} < R_{c} < R_{out}$ and the clumpy structure is the combination of the inner disc extending from $R_{in}$ to $R_{c}$ and the outer cloudy torus (wind region) extending from $R_{c}$ to $R_{out}$ (see the second panel in Figure 1).

\textbf{Class C}: $R_{c}>R_{out}$ and the clumpy structure is disc-like (see the third panel in Figure 1).

Depending on the observational data which we know about each AGN we suggest three methods to find what category it belong to.

1) For some AGNs with black hole mass independently estimated from the $M - \sigma$ relationship (e.g., \citealp{Onken04,Woo10,Graham11,Grier13}), we can determine the Eddington luminosity defined by $ L_{Edd}=1.3 \times 10^{38} M/M_{\odot} erg/s $. On the other hand, by measuring the bolometric luminosity, $L$, we have the Eddington ratio $ l=L/L_{Edd}$. Also, we assume $ 1.2 \times 10^{21}cm^{-2}< N_{0} < 1.5 \times 10^{24}cm^{-2}$ for the column density (e.g., \citealt{Marconi08}) and $1 < s < 2.5$ (e.g., \citealt{Rees89}). We can then calculate the value of $R_{c}$ from equation (\ref{eq11}) and compare it with $R_{in}$ and $R_{out}$ specified by reverberation mapping technique (e.g., \citealt{Krolik91}) to determine what class each object belongs to.

2) For cases for which reverberation mapping is not available, the $R-L$ relationship (\citealp{Kaspi00,Bentz06}) can help us to estimate the BLR radius and compare it with $R_{c}$ estimated by the method described above.

3) For AGNs with unknown black hole mass, if we assume $R_{in}=10R_{Sch}$ and $R_{out}=1000R_{Sch}$ we can define the parameter $\chi$ as $\chi =R_{c}/R_{out}= (0.005c^{2}R_{0}/GM)(2\mu \sigma_{T}N_{0}/3l)^{3/2s}$. Clearly $ \chi < 0.01 $ implies $R_{c} < R_{in}$ and such systems belong to class A. Also, in the cases for which $0.01 < \chi < 1$ and $\chi > 1$, we see that they belong to the classes B and C respectively. Assuming that $s=3/2$ we can combine the black hole mass with the Eddington ratio to get $\chi$ as
\begin{equation}\label{eq19}
\chi = 4.89 \times 10^{21} \frac{\mu N_{0}}{L}.
\end{equation}
The parameter $\chi$ derived from equation (\ref{eq19}) can give the class of BLR structure for AGNs with unknown black hole mass. The top and bottom panels in Figure \ref{figure2} show $\log \chi$ as a function of $\log L$ and $\log N_{0}$ respectively. In Figure 2 we can see that for quasars with luminosities ranging from $10^{44} erg/s $ to $ 10^{47} erg/s$, we have $ -2<\log \chi<0 $ for less luminous systems and $\log \chi<-2 $ for the brightest cases. We therefore expect that the BLR structure of AGNs is similar to classes A and B for the higher and lower luminosity cases. For Seyfert galaxies with $ 10^{41} erg/s < L < 10^{44} erg/s$, the  expected distribution of clouds is as in classes B and C for higher and lower luminosity objects.  Finally for all the LLAGNs with $ L<10^{41} erg/s$ we see $\log \chi$ is positive and BLR structure is as in class C (i.e., disc-like). Some studies, through a discussion on the shape of broad emission lines, have shown the existence of disc sturcture (class C) in LLAGNs (e.g., \citealp{Eracleous03,Storchi16}). Moreover they have found the existence of a non-disc region (outer torus in class B) in outer parts of BLR in Seyfert galaxies. 

As the final step in this section, we solve equations (\ref{eq15}) and (\ref{eq18}) for the disc-like approximation to obtain the volume number density $n$. Because of the assumption of strong coupling between the BLR clouds and the hot intercloud medium we have $\langle v_{R} \rangle =w_{R}$. Furthermore, if we assume an advection-dominated accretion flow (ADAF) for the model describing the intercloud gas and because of the self-similar solution we can write $ w_{R}=-\alpha c_{1}v_{k,mid}$ (e.g., \citealt{Narayan94}), where $\alpha $ and $ c_{1}$ are two constants in order of unity and $ v_{k,mid}$ is the Keplerian velocity in the midplane. On the other hand, from equation (\ref{eq15}) we can see that $ nR\langle v_{R} \rangle $ does not depend on $R$, so
\begin{equation}\label{eq20}
n(R,z)=-\frac{\Lambda(z)}{\alpha c_{1}\sqrt{GM}}R^{-\frac{1}{2}},
\end{equation}
where $\Lambda(z)$ give the vertical dependence of $n$. By substituting equation (\ref{eq20}) into equation (\ref{eq18}) and after some algebraic manipulations, equation (\ref{eq18}) can be written as
\begin{equation}\label{eq21}
\langle v^{2}_{z} \rangle \frac{\partial \Lambda(z)}{\partial z} +\Omega_{k,mid}^{2}z\left[1-\left(\frac{R}{R_{c}}\right)^{2s/3}\right]\Lambda(z)=0.
\end{equation}
If we substitute $\Lambda(z)$ derived by integrating equation (\ref{eq21}) into equation (\ref{eq20}), the volume number density is given by
\begin{equation}\label{eq22}
n(R,z)=-\frac{k_{0}}{\alpha c_{1}\sqrt{GM}}R^{-\frac{1}{2}}\exp \left(-\frac{z^{2}}{2h^{2}_{g}}\right),
\end{equation}
where $k_{0}$ is the constant of integration and $h_{g}$ is the scale height which is
\begin{equation}\label{eq23}
h_{g}(R)=\frac{\sigma_{z}}{\Omega_{k,mid}}\sqrt{\frac{1}{1-(R/R_{c})^{\frac{2s}{3}}}}.
\end{equation}
Integrating $n$ over all positions occupied by the clouds, we calculate $ k_{0}$ in terms of the total number of the cloud $ n_{tot}$, and substitute it into equation (\ref{eq22}) to derive $n$ as
\begin{equation}\label{eq24}
n(R,z)=\frac{\sqrt{GM}}{(2\pi)^{3/2}}\frac{n_{tot}}{\sigma_{z}\gamma}R^{-1/2}\exp \left(-\frac{z^{2}}{2h^{2}_{g}}\right),
\end{equation}
where $\sigma_{z}=\sqrt{\langle v^{2}_{z} \rangle}$ is $z$-component of the velocity dispersion and $\gamma $ is defined by $\gamma=\int_{R_{in}}^{R_{out}}\sqrt{1/1-(R/R_{c})^{\frac{2s}{3}}}R^{2}dR$, and where $R_{in}$ and $R_{out}$ are assumed to be 1 and 10 light day respectively. To calculate the surface number density, $\Sigma$, we integrate equation (\ref{eq24}) over $z$ to get
\begin{equation}\label{eq25}
\Sigma (R)=\frac{1}{2\pi}\frac{n_{tot}}{\gamma}R\sqrt{\frac{1}{1-(R/R_{c})^{\frac{2s}{3}}}}.
\end{equation}
In all subsequent sections, we adopt $\mu =0.61$.

\section{VIRIAL FACTOR}\label{s3}

\subsection{Disc-like configurations}\label{ss31}
In the last section we saw that in class C we have a disc-like distribution for the BLR. There are now two questions: what is the geometric shape of the clumpy disc? and is its thickness small compared to its radial sizes? In this section, we will answer to these questions and explore the role of various physical parameters on the shape and thickness of the clumpy disc. According to equation (\ref{eq24}) we consider the curved surface $ z=h_{g}(R)$ as the boundary between the clumpy and non-clumpy regions. Through plotting the scale height profile as a function of the radial distance in Figure \ref{figure3} we show the geometric shape of the clumpy disc.

In Figure \ref{figure3}, we see that there is a positive correlation between the radial distance and scale height. Panels a and b show the role of the central luminosity and black hole mass on the thickness of the clumpy disc respectively. From these we see that with an increase in the central luminosity, the disc thickness increases, and that an increase in the black hole mass leads to a decrease in thickness. This is because as $l$ increases, radiative pressure dominates and pushes the clouds away from the midplane. However, an increase in the black hole mass leads to an increase in the gravitational attraction pulling the clouds toward the midplane. Panels c and d show that there is similar behaviour for $h_{g}$ as a function of $\sigma_{z}$ and $s$. With increases in both of them, $h_{g}$ increases as well. Figure (\ref{figure3}) also shows that, as suggested by some authors (e.g., \citealp{Dumont90b,Goad12}), the clumpy disc is flared (bowl-shaped). The thickness of the clumpy disc is small compared to the radial size ($ h_{g} \approx 0.1R$). This is important.

Obviously, as the central luminosity declines, the clouds exposed to this radiation reprocess less amount of energy. As a result the broad emission lines are weak in LLAGNs. However, in addition to this, the small thickness of the clumpy disc results in a small solid angle being covered by the clumpy disc and this leads to little capture of the central radiation. The fraction of radiation captured is given by $ d\Omega /4\pi \approx \Theta^{2}/2 \approx h_{g}^{2}/2R^{2}$ which is of the order of $ 10^{-3}$, where $ d\Omega $ is the solid angle covered by the clumpy disc and $\Theta \approx h_{g}/R$. Therefore, all the clouds located in the BLR of LLAGN can only receive $ 0.001 $ of the central radiation which itself is in order of $ 10^{-4}-10^{-5}$ the Eddington luminosity and this leads to the presence of very weak broad emission lines in the spectra of LLAGNs. In the very lowest luminosity cases ($ L=10^{-8} - 10^{-9} L_{Edd}$), the small thickness of the clumpy disc together with the faintness of the central source can cause that the broad emission lines to fall below observational detection thresholds and this can explain the lack of detection of any broad emission lines in such faint objects.

Figures \ref{figure4}, \ref{figure5} and \ref{figure6} respectively clarify the effect of $l$, $s$ and $n_{tot}$ on the surface number density $\Sigma $ and volume number density $n$ which are plotted versus the radius $R$. In these three figures, the first panels present the behaviour of $\Sigma $ versus $R$ and the second panels show the variations of $n$ with radius at $ z=0, 0.2 $ and $ 0.5 $ light day (lt-d). Values of the fixed parameters are specified in the captions. In these figures, the profiles of the surface number density show that most of the clouds are located in the outer parts of the BLR. However the profiles of the volume number density are somewhat different. Whilst in the midplane most of the clouds are close to the central black hole, at $ z=0.2$ lt-d, the value of $n$ in the central parts is almost zero and with increasing the radius, it increases to a maximum and then, for larger radial distances $n$ gradually declines. At $ z = 0.5$ lt-d the behaviour of the curve is similar to that at $z = 0.2$ lt-d but the peak of the curve moves towards radii. Note that, by considering the positive correlation between $ h_{g}$ and $R$, there is no contradiction between the behaviour of $\Sigma $ and $n$ as a function of $R$.

Panel 4a shows that as $l$ increases, the ratio of the clouds in the outer regions of the BLR to those in the inner regions increases as well. On the other hand, by assuming strong coupling between the clouds and the ambient medium, we have $\langle v_{\phi} \rangle=w_{\phi} \propto v_{k,mid}$ and $\langle v_{R} \rangle=w_{R} \propto v_{k,mid}$. Consequently, we conclude that, with increasing of $l$, the number of slowly moving clouds in the outer parts increases and the number of quickly-moving clouds in the inner parts is reduced. In other words, as obtained by \citet{Netzer10}, we expect that with increasing $l$, the width of broad emission lines $\textsc{FWHM}$ decreases. Panel 4b shows that an increase in $l$ leads to the reduction in the number of the clouds near the midplane. This is because they are distributed to higher altitudes. As is shown in Figure \ref{figure5} the effect of the density index is similar to that of the central luminosity. Finally, in Figure \ref{figure6} we see that an increase in the total number of the clouds leads, as is expected, to an increase in the value of $\Sigma $ and $n$.

\begin{figure}
\centering
\includegraphics[width=0.79\columnwidth]{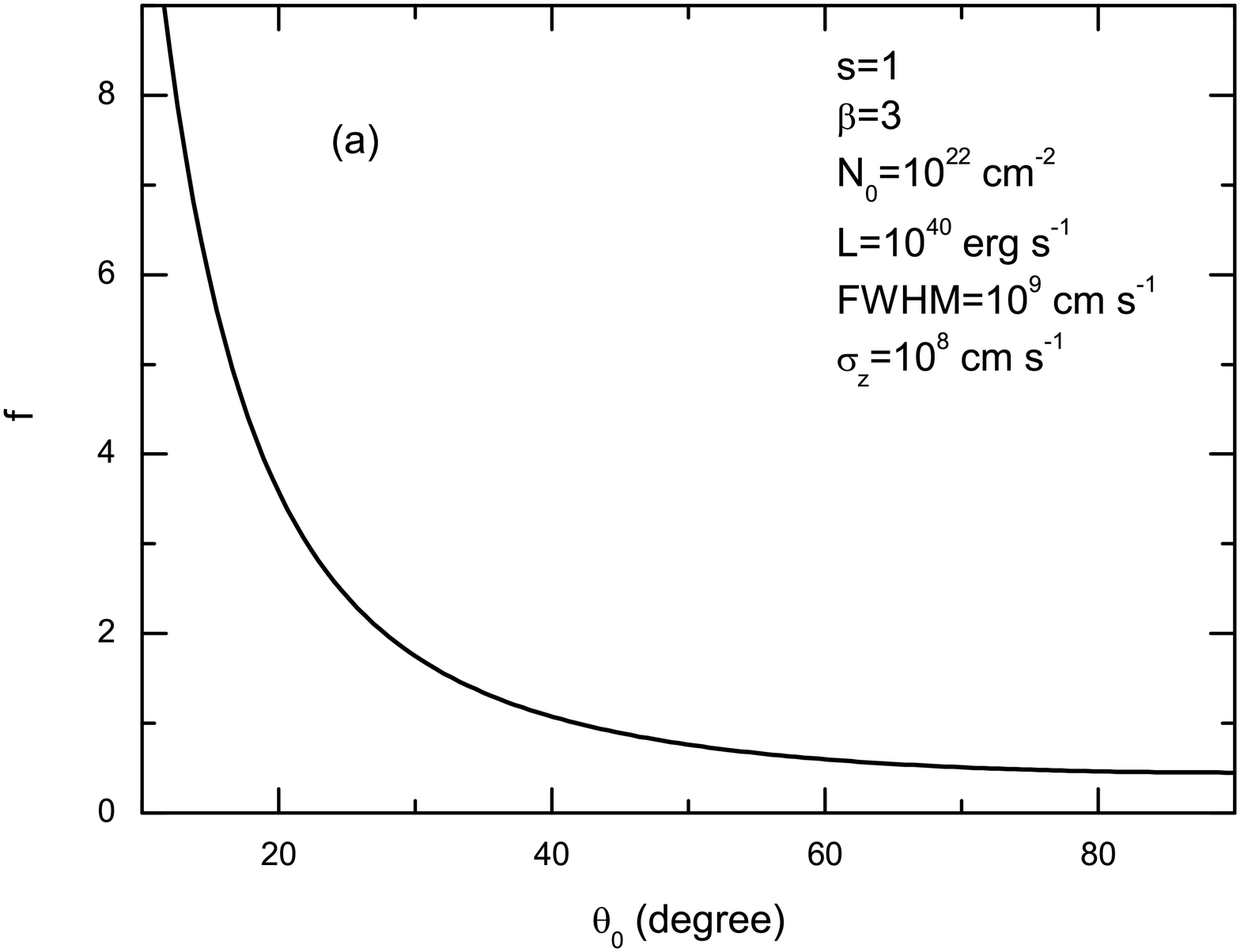}
\includegraphics[width=0.79\columnwidth]{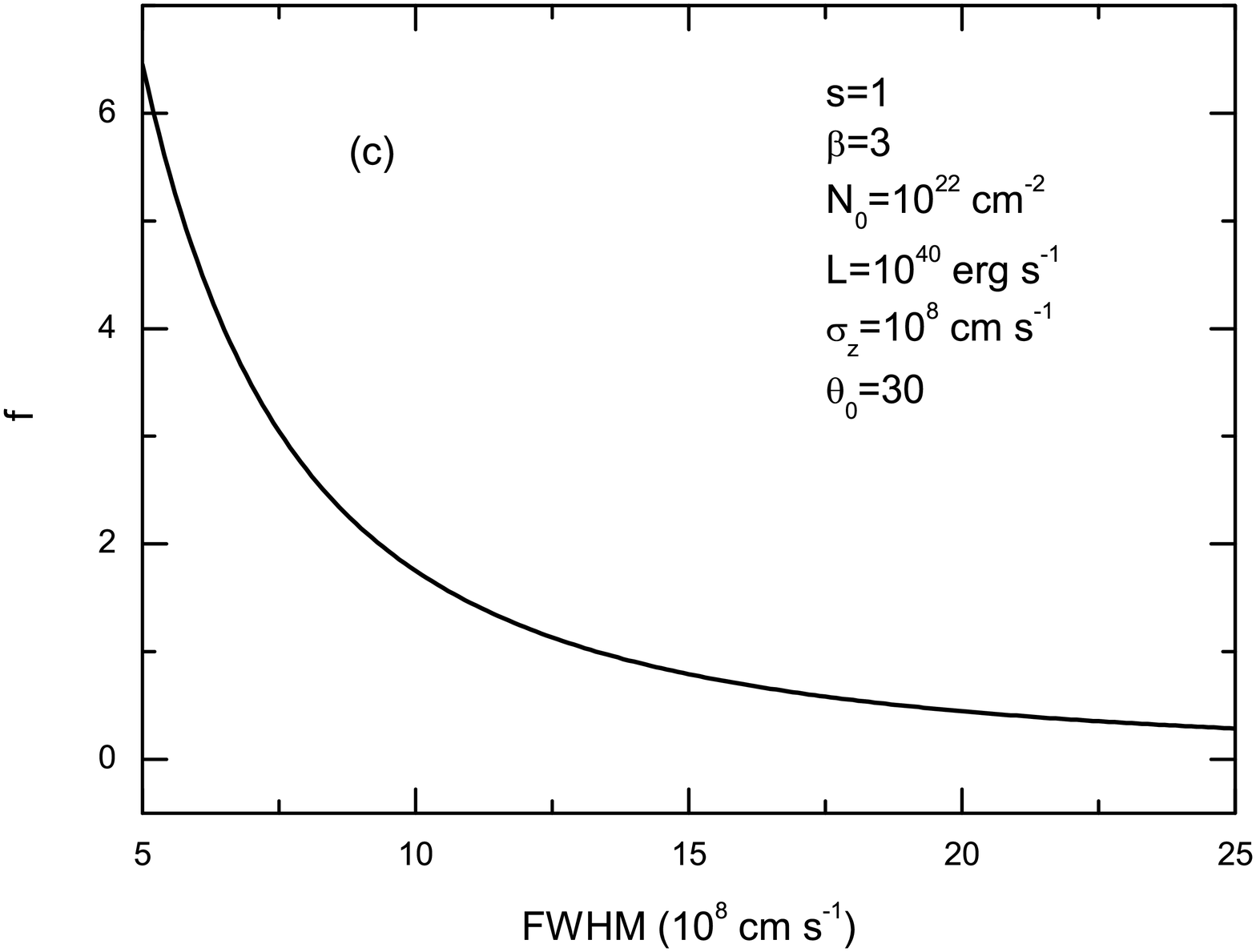}
\includegraphics[width=0.79\columnwidth]{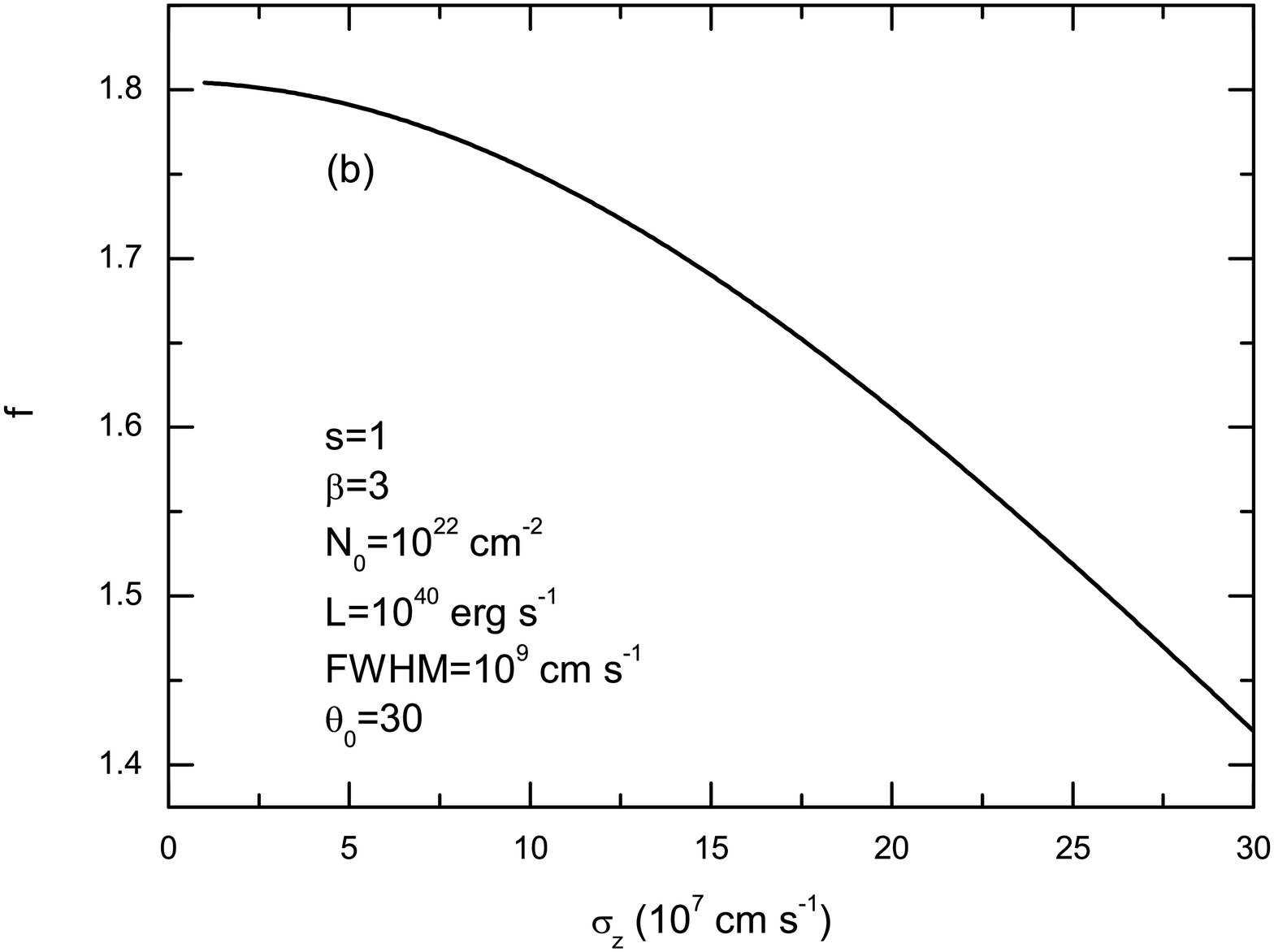}
\caption{The panels (a), (b) and (c) show the virial factor as a function of inclination angle $\theta_{0}$, z-component of the velocity dispersion $\sigma_{z}$, and width of broad emission line $\textsc{FWHM}$ respectively. Values of other fixed parameter are listed on each panel.}
\label{figure7}
\end{figure}
\begin{figure}
\centering
\includegraphics[width=0.79\columnwidth]{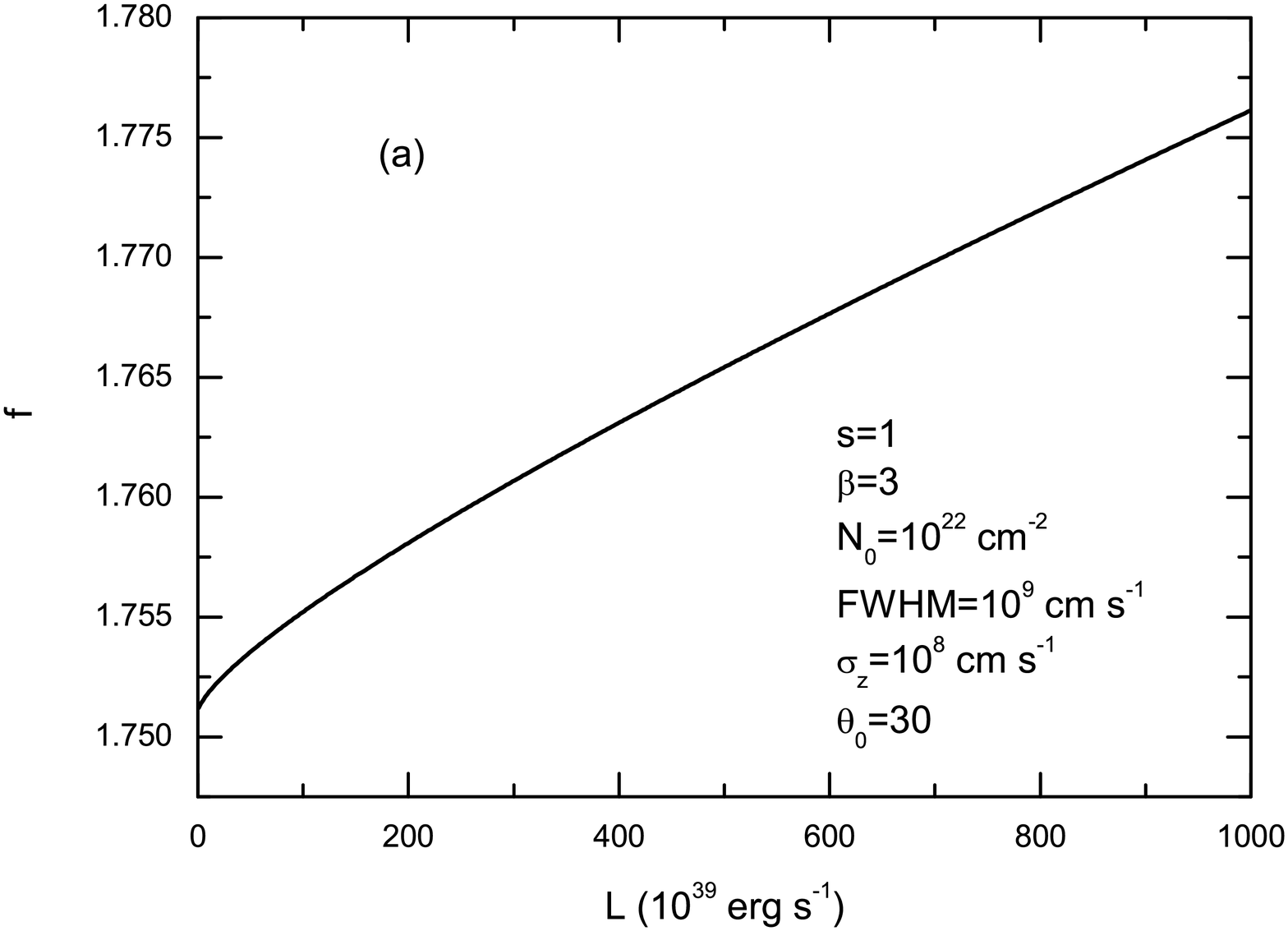}
\includegraphics[width=0.79\columnwidth]{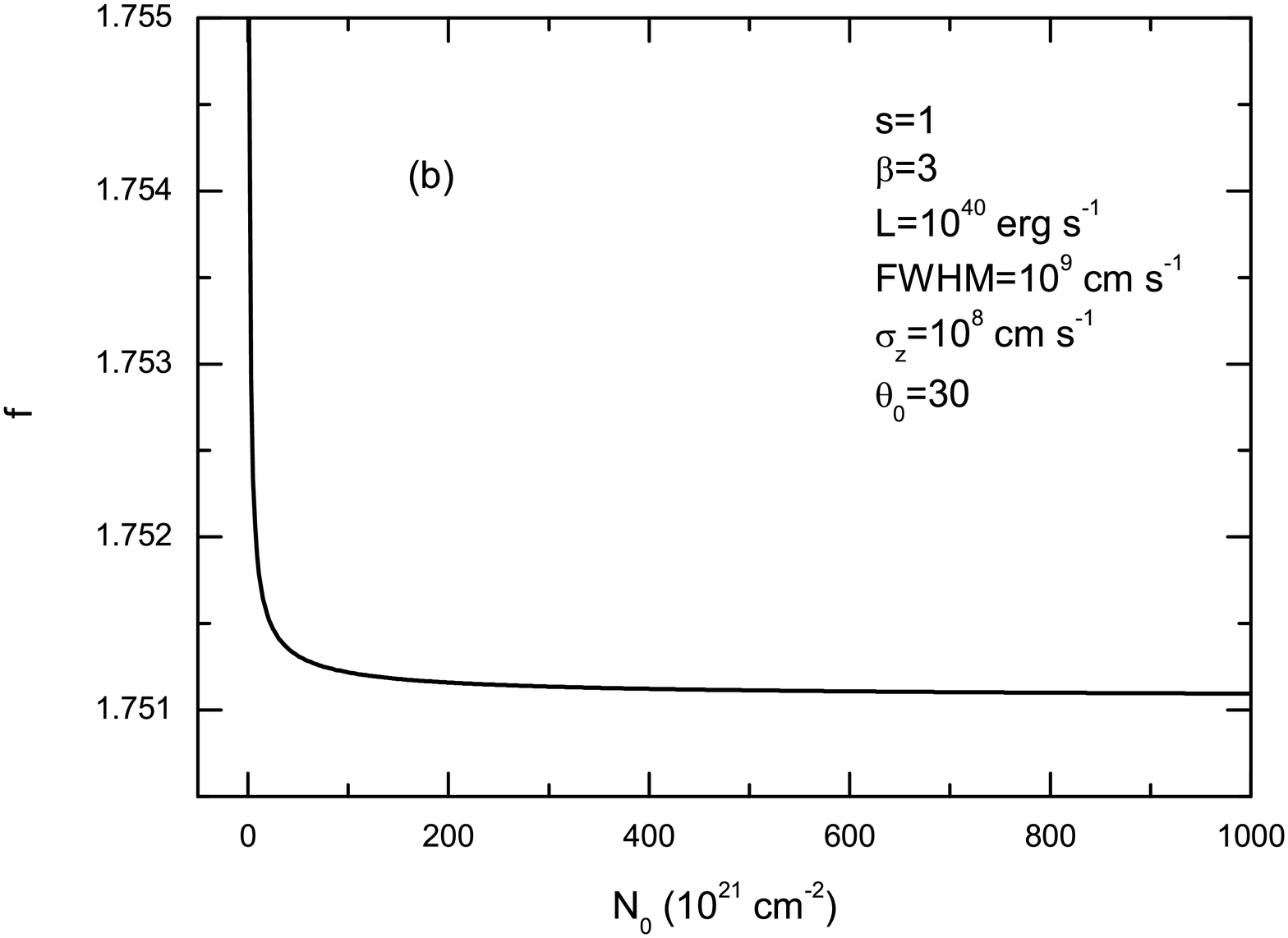}
\includegraphics[width=0.79\columnwidth]{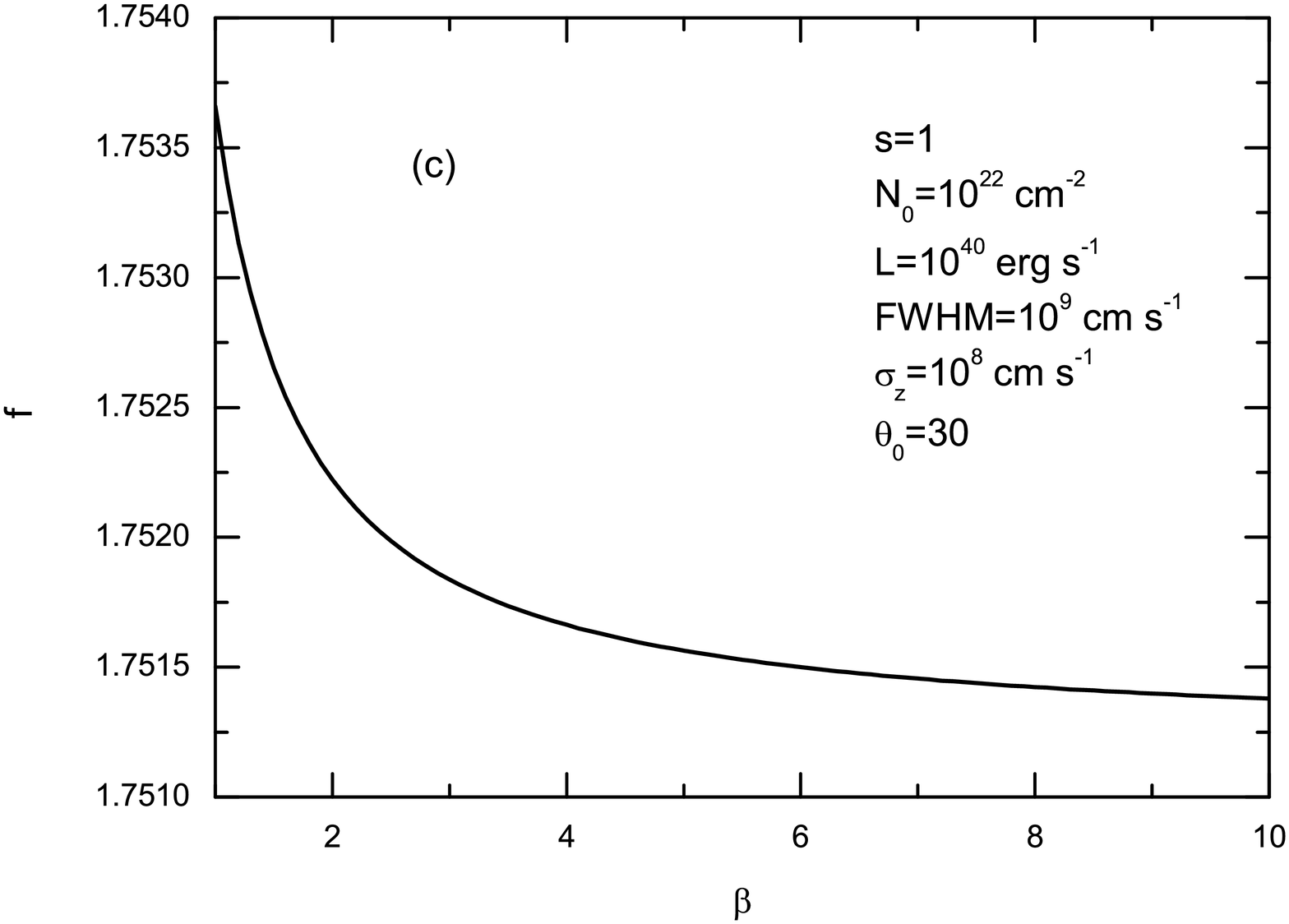}
\includegraphics[width=0.79\columnwidth]{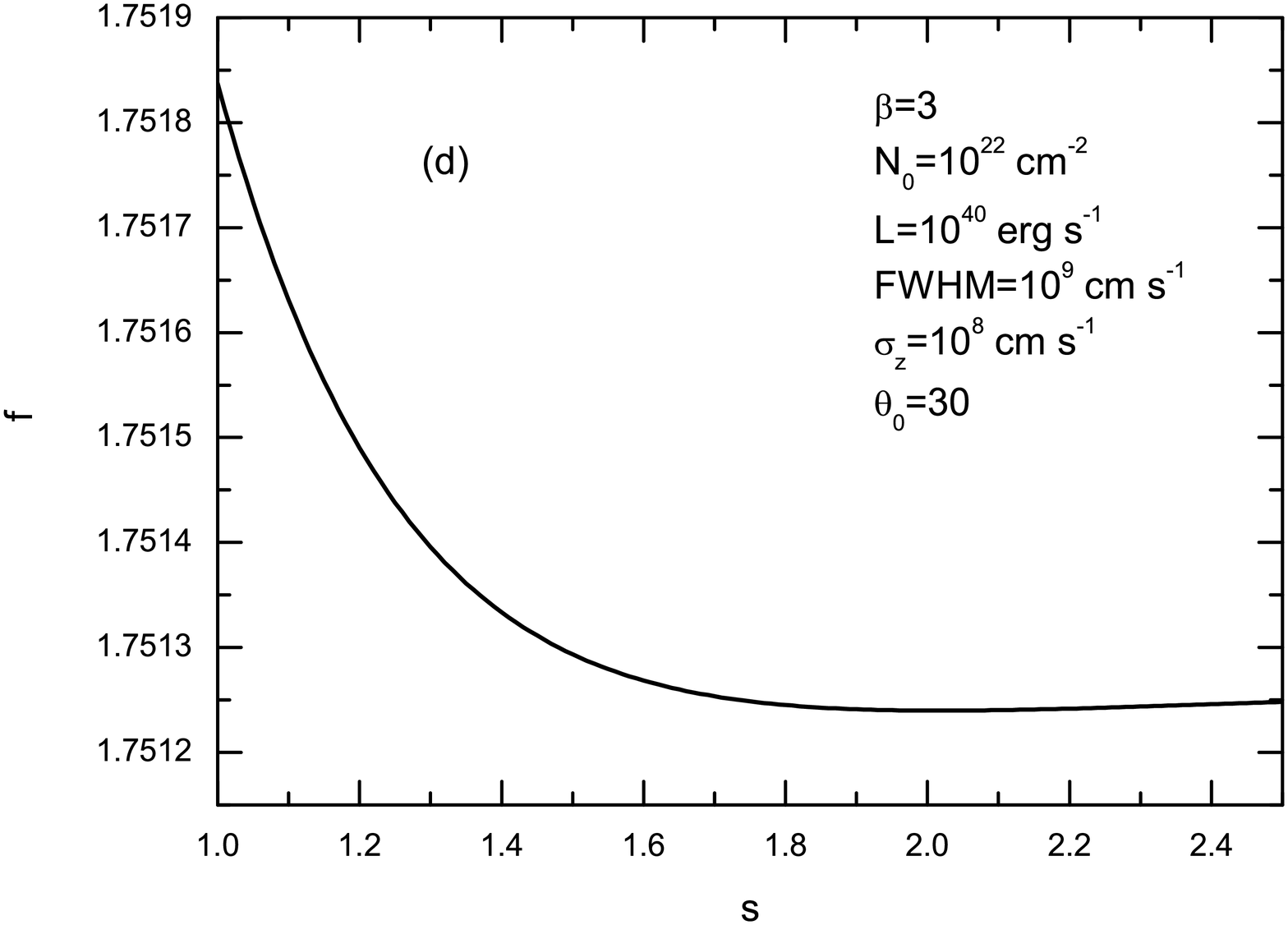}
\caption{The panels (a), (b), (c) and (d) show the virial factor as a function of bolometric luminosity ($L$), column density ($N_{0}$), $\beta$ and the density index ($s$) respectively. Values of other fixed parameter are listed on each panel.}
\label{figure8}
\end{figure}

\subsection{Calculation of Virial factor for disc-like structure}\label{ss32}

We can derive the virial factor, $f$, for the disc-like clumpy structure as a function of the kinematic parameters of the BLR and the inclination angle (the angle between the observer's line of sight and the axis of symmetry of the thin-disc. In Appendix \ref{a2}, we will show that the averaged line of sight velocity square $\langle v_{n}^{2} \rangle _{avr}$ can be written as
\begin{equation}\label{eq26}
\langle v_{n}^{2} \rangle _{avr}=\frac{1+\beta}{2}\langle v_{R}^{2} \rangle \sin^{2} \theta_{0}+\langle v_{z}^{2} \rangle \cos^{2} \theta_{0},
\end{equation}
where $\theta_{0}$ is the inclination angle and $\beta = \langle v_{\phi}^{2} \rangle/\langle v_{R}^{2} \rangle $ is taken to be constant. As in section (\ref{s2}), if we assume $\langle v_{z} \rangle =0$, we have $\langle v_{z}^{2} \rangle=\sigma_{z}^{2}$ which is taken to be constant. However, $\langle v_{R}^{2} \rangle $ has to be derived by solving equation (\ref{eq16}). Dividing equation (\ref{eq16}) into $ n$, we can write
\begin{equation}\label{eq27}
\frac{\partial \langle v^{2}_{R} \rangle}{\partial R}+\left[\frac{\partial \ln (n)}{\partial R}+\frac{1-\beta}{R}\right]\langle v^{2}_{R}\rangle =-\Omega_{k,mid}^{2}R\left[1-\left(\frac{R}{R_{c}}\right)^{\frac{2s}{3}}\right],
\end{equation}
where $\partial \ln (n)/\partial R $ is derived by the substitution of $ n(R,z)$, from equation (\ref{eq24}). Equation (\ref{eq27}), as a first order differential equation, then becomes
\begin{equation}\label{eq28}
\frac{\partial \langle v^{2}_{R} \rangle}{\partial R}+\left[\frac{1-2\beta}{2R}+\frac{z^{2}}{h_{g}^{3}}\frac{dh_{g}}{dR}\right]\langle v^{2}_{R}\rangle =-\Omega_{k,mid}^{2}R\left[1-\left(\frac{R}{R_{c}}\right)^{\frac{2s}{3}}\right] .
\end{equation}
Finding the integrating factor, we can write the solution of equation (\ref{eq28}) as
\[\langle v^{2}_{R} \rangle =-GMR^{\frac{2\beta -1}{2}}\exp\left(\frac{z^{2}}{2h_{g}^{2}}\right)\int R^{-\frac{3+2\beta}{2}}\left[1-\left(\frac{R}{R_{c}}\right)^{\frac{2s}{3}}\right]\]
\begin{equation}\label{eq29}
\times \exp\left(-\frac{z^{2}}{2h_{g}^{2}}\right)dR.
\end{equation}
In this integral, the range of the variation of $\exp(-z^{2}/2h_{g}^{2})$ as a function of $R$, is given by $ [\partial \exp(-z^{2}/2h_{g}^{2})/\partial R]\Delta R \approx \exp(-z^{2}/2h_{g}^{2})(z^{2}/h_{g}^{3})(h_{g}/R)\Delta R $ which is in the order of unity. This is because, in thin disc, we have $ z \approx h_{g}$ and $\Delta R \approx R $. On the other hand, by assuming $R_{out}\approx 10R_{in}$, $\beta=3 $ and $ s=3/2$, we see that, as $R$ increases from $R_{in}$ to $R_{out}$, the value of the other terms inside the integral, $R^{-(3+2\beta)/2}[1-(R/R_{c})^{2s/3}]$, becomes smaller by a factor of 1000. We therefore assume that the value of $\exp(-z^{2}/2h_{g}^{2})$ remains constant and by taking it out of the integral, $\langle v_{R}^{2} \rangle $ is given by
\begin{equation}\label{eq30}
\langle v_{R}^{2}\rangle = \frac{GM}{R}\left[\frac{2}{1+2\beta}+\frac{6}{4s-6\beta -3}\left(\frac{R}{R_{c}}\right)^{\frac{2s}{3}}\right]+c_{0},
\end{equation}
where $ c_{0}$ is the constant of integration calculated as follows. As we discussed in the subsection (\ref{ss22}), for the thin-disc structure, we have $R<R_{c}$. As a result, in order to find the value of $ c_{0}$, we suppose  that $\langle v_{R}^{2} \rangle + \langle v_{\phi}^{2} \rangle = (1+\beta)\langle v_{R}^{2} \rangle \approx 2\textsc{FWHM}^{2}$ at $R=0.5R_{c}$. Finally, by substituting the constant of the integration into equation (\ref{eq30}), $\langle v_{R}^{2} \rangle $ can be expressed as
\[\langle v_{R}^{2}\rangle = \frac{GM}{R}\left\{\left[\frac{2}{1+2\beta}+\frac{6}{4s-6\beta -3}\left(\frac{R}{R_{c}}\right)^{2s/3}\right]-\left[\frac{4}{1+2\beta}\right.\right.\]
\begin{equation}\label{eq31}
\left. \left. +\frac{12}{4s-6\beta -3}\left(\frac{1}{2}\right)^{2s/3}\left(\frac{R}{R_{c}}\right)\right]+\frac{2}{1+\beta}\frac{R}{GM}\textsc{FWHM}^{2}\right\}.
\end{equation}

By substituting $\langle v_{R}^{2}\rangle $ into equation (\ref{eq26}), the virial factor defined by $ f=GM/R\langle v_{n}^{2} \rangle _{avr}$, is given by
\[f=\left\{\left[\left(\frac{1+\beta}{1+2\beta}+\frac{3+3\beta}{4s-6\beta -3}\left(\frac{R}{R_{c}}\right)^{2s/3}\right)-\left(\frac{2+2\beta}{1+2\beta}\right.\right.\right.\]
\[\left.\left.+\frac{6+6\beta}{4s-6\beta -3}\left(\frac{1}{2}\right)^{2s/3}\left(\frac{R}{R_{c}}\right)\right)+\frac{R}{GM}\textsc{FWHM}^{2}\right]\sin^{2}\theta_{0}\]
\begin{equation}\label{eq32}
\left.+\frac{R}{GM}\sigma_{z}^{2}\cos^{2}\theta_{0}\right\}^{-1} .
\end{equation}
Finally, assuming $R \approx R_{out} \approx 1000R_{Sch}$, the virial factor becomes
\[f=\left\{\left[\left(\frac{1+\beta}{1+2\beta}+\frac{3+3\beta}{4s-6\beta -3}\right)\chi^{-\frac{2s}{3}}-\left(\frac{2+2\beta}{1+2\beta}\right.\right.\right.\]
\[\left.\left.\left.\left.+\frac{6+6\beta}{4s-6\beta -3}\left(\frac{1}{2}\right)^{\frac{2s}{3}}\right)\chi^{-1}+2000\left(\frac{\textsc{FWHM}}{c}\right)^{2}\right]\sin^{2}\theta_{0}\right.\right.\]
\begin{equation}\label{eq33}
\left.+2000\left(\frac{\sigma_{z}}{c}\right)^{2}\cos^{2}\theta_{0}\right\}^{-1},
\end{equation}
where $\chi = R_{c}/R_{out}$ is defined by equation (\ref{eq19}).

Figures (\ref{figure7}) and (\ref{figure8}) show the variation of the virial factor as a function of the various parameters. In spite of all the approximations we have used, we see that the value of $f$ is of the order of unity.  This is in an agreement with previous works (e.g., \citealp{Onken04,Woo10,Graham11,Grier13}). Figure (\ref{figure7}) shows that the virial factor changes significantly with the inclination angle $\theta_{0}$, the width of broad emission lines $\textsc{FWHM}$, and $z$-component of the velocity dispersion $\sigma_{z}$. From the first panel of Figure (\ref{figure7}), it can be seen that as $\theta_{0}$ increases from 0 to 40 (type1 AGNs), the value of $f$ rapidly falls from nearly 9.0 to 1.0 and with increasing $\theta_{0}$ from 40 to 90 (type2 AGNs), it gradually decreases from nearly 1.0 to 0.5. The negative correlation between $ f $ and $ \theta_{0} $ is similar to the finding derived by \citet{Pancoast14} for five Seyfert galaxies including: Arp 151, Mrk 1310, NGC~5548, NGC~6814 and SBS 1116+583A. In the second panel of Figure (\ref{figure7}) showing similar behaviour for the virial factor as a function of $\textsc{FWHM}$, we see $ 0.5 \lesssim f \lesssim 6.5 $. The anticorrelation between $f$ and $\textsc{FWHM}$ has been confirmed by \citet{Brotherton15}. Finally the third panel shows that the value of $ f $ is nearly between 1.4 to 1.8. However, unlike two first panels, the slope of the curve in $ f - \sigma_{z} $ diagram is shallow for lower values of $ \sigma_{z} $ and steep for higher values.

From Figure (\ref{figure8}) we see that with increasing $\beta$, $s$ and $N_{0}$ the value of the Virial factor, $ f $, decreases and with increasing $L$ it increases. But we have to note that the range of variation of $ f $ is so small (of the order of 0.01 for the variation of $ L $ and 0.001 for the variation of three other quantities). In other words, $f$ is relatively insensitive to the variation of $L$, $\beta$, $s$ and $N_{0}$. The insensitive correlation between the Virial factor and bolometric luminosity is in agreement with the results found by \citet{Netzer10}. They found that while cloud orbits are strongly affected by radiation pressure, there is a relatively small change in $r_{BLR}\textsc{FWHM}^{2}$.  This means that radiation pressure does not change the value of virial factor significantly. 

\section{CONCLUSIONS}\label{s5}

In this work, considering the clouds as a collisionless ensemble of particles, we employed the cylindrical form of Jeans equations calculated in section 2 to describe a geometric model for their distributions in the BLR. The effective forces in this study are the Newtonian gravity of the black hole, the isotropic radiative force arisen from the central source and the drag force for linear regime. Taking them into account we showed that there are three classes for BLR configuration: (A) non disc (B) disc-wind (C) pure disc structure (see Figure \ref{figure1}). We also found that the distribution of BLR clouds in the brightest quasars belongs to class A, in the dimmer quasars and brighter Seyfert galaxies it belongs to class B, and in the fainter Seyfert galaxies and all LLAGNs (LINERs) it belongs to class C. 

Then we derived the Virial factor, $ f $, for disc-like structures and found a negative correlation for $f$ as a function of the inclination angle, the width of broad emission line and z-component of the velocity dispersion. We also found $ 1.0 \lesssim f \lesssim 9.0 $ for type1 AGNs and $ 0.5 \lesssim f \lesssim 1.0 $ for type2 AGNs. Moreover we saw that $ f $ approximately varies from 0.5 to 6.5  for different values of $\textsc{FWHM}$ and from 1.4 to 1.8  for different values of $ \sigma_{z} $. We also indicated that $ f $ doesn't change significantly with the variations of bolometric luminosity, column density of each cloud, density index and $ \beta = \langle v_{\phi}^{2} \rangle/\langle v_{R}^{2} \rangle $ and the maximum change in the value of $ f $ is of order of 0.01.

In introduction, we mentioned that since each group take a different sample of AGNs, they find different values for average virial factor, $ \langle f \rangle $. However different values leads to significant uncertainties in the estimation of black hole mass. On the other hand, in this paper, we saw that $ f $ significantly changes with the inclination angle $ \theta_{0} $ and $ \textsc{FWHM} $ (Figure \ref{figure7}). Therefore in order to have more accurate estimation for black hole mass, we suggest observational campaigns to divide a sample of objects into a few subsamples based on the value of $ \theta_{0} $ and $ \textsc{FWHM} $ of objects and then determine the value of $ \langle f \rangle $ for each subsample separately. Therefore we will have several values for $ \langle f \rangle $. Finally regarding the value of $ \theta_{0} $ and $ \textsc{FWHM} $ of each object with unknown black hole mass, we use the appropriate value of $ \langle f \rangle $ in the virial theorem to have more accurate estimation of black hole mass.

\section*{Acknowledgements}
I am very grateful to the referee, Jian-Min Wang, for his very useful comments which improved the manuscript. I also thank Scott Tremaine for his useful suggestions that clarified some points about the extended form of the collisionless Boltzmann equation.

\newpage
\appendix
\section{DERIVATION OF JEANS EQUATION}\label{a1}
In this appendix, we derive the Jeans equations from the collisionless Boltzmann equation (CBE) for the particles which their movements are affected by both position-dependent and velocity dependent forces. First we demonstrate the mathematical formula which is used in this way several times. From the Product Rule for Derivatives, we have
\begin{equation}\label{eqa1}
\int \psi \frac{\partial F}{\partial v_{i}}d^{3}v=\int \frac{\partial (\psi F)}{\partial v_{i}}d^{3}v - \int F\frac{\partial \psi}{\partial v_{i}}d^{3}v,
\end{equation}
where, $\psi$, and $F$ are an arbitrary function and the distribution function respectively. Moreover $ v_{i}$ represents the velocity components in the cylindrical coordinates ($ v_{R}, v_{\phi}, v_{z}$). Since we don't have any particle with infinite velocity, so the value of $F$ for sufficiently large velocities is equal to zero (e.g., \citealt{Binney87}) and, due to the divergence theorem, the first term on the right side of equation (\ref{eqa1}) vanishes and we have
\begin{equation}\label{eqa2}
\int \psi \frac{\partial F}{\partial v_{i}}d^{3}v=-\int F\frac{\partial \psi}{\partial v_{i}}d^{3}v.
\end{equation}
On the other hand, we define the averaged parameter in the velocity space as $\langle X \rangle=n^{-1} \int X F d^{3}v $ where $ X $ is an arbitrary parameter and $n$ is the volume number density in position-place which is calculated by $ n=\int F d^{3}v $. Integrating CBE in the velocity space, we have 
\[\int \frac{\partial F}{\partial t}d^{3}v+\int v_{R}\frac{\partial F}{\partial R}d^{3}v+\int \frac{v_{\phi}}{R}\frac{\partial F}{\partial \phi}d^{3}v+\int v_{z}\frac{\partial F}{\partial z}d^{3}v\]
\[+\int \left(a_{R}+\frac{v_{\phi}^{2}}{R}\right)\frac{\partial F}{\partial v_{R}}d^{3}v+\int \left(a_{\phi}-\frac{v_{R}v_{\phi}}{R}\right)\frac{\partial F}{\partial v_{\phi}}d^{3}v\]
\begin{equation}\label{eqa3}
+\int a_{z}\frac{\partial F}{\partial v_{z}}d^{3}v+\int F\frac{\partial a_{R}}{\partial v_{R}}d^{3}v+\int F\frac{\partial a_{\phi}}{\partial v_{\phi}}d^{3}v+\int F\frac{\partial a_{z}}{\partial v_{z}}d^{3}v=0,
\end{equation}
since the velocity components ($ v_{R}, v_{\phi}, v_{z}$) don't depend on $R$, $\phi $ and $z$ and also the range of velocities over which we integrate depends on neither time nor space, so the partial derivatives respect to both time ($\partial /\partial t$) and space ($\partial /\partial R, \partial /\partial \phi, \partial /\partial z$) in the first four terms on the left side of equation (\ref{eqa3}) are taken outside. Also by using equation (\ref{eqa2}) for the fifth, sixth and seventh terms of equation (\ref{eqa3}), the first relation of the Jeans equations is gained as
\begin{equation}\label{eqa4}
\frac{\partial n}{\partial t} + \frac{1}{R} \frac{\partial}{\partial R} (Rn\langle v_{R} \rangle)+\frac{1}{R}\frac{\partial}{\partial \phi} (n\langle v_{\phi} \rangle)+\frac{\partial}{\partial z} (n\langle v_{z} \rangle)=0,
\end{equation}
the derivation of other relations of the Jeans equations is similar to that of the first one. By multiplying CBE by $ v_{R} , v_{\phi}$ and $ v_{z}$ respectively and integrating them in the velocity-space, other equations can be given by
\[\frac{\partial}{\partial t} (n\langle v_{R} \rangle)+\frac{\partial}{\partial R} (n\langle v^{2}_{R} \rangle)+\frac{1}{R}\frac{\partial}{\partial \phi} (n\langle v_{R}v_{\phi} \rangle)+\frac{\partial}{\partial z} (n\langle v_{R}v_{z} \rangle)\]
\begin{equation}\label{eqa5}
+n\frac{\langle v^{2}_{R}\rangle -\langle v^{2}_{\phi}\rangle}{R}-n \langle a_{R} \rangle =0,
\end{equation}
and
\[\frac{\partial}{\partial t} (n\langle v_{\phi} \rangle)+\frac{\partial}{\partial R} (n\langle v_{R}v_{\phi} \rangle)+\frac{1}{R} \frac{\partial}{\partial \phi} (n\langle v^{2}_{\phi} \rangle)+\frac{\partial}{\partial z} (n\langle v_{\phi}v_{z} \rangle)\]
\begin{equation}\label{eqa6}
+\frac{2n}{R}\langle v_{\phi}v_{R}\rangle -n \langle a_{\phi} \rangle =0,
\end{equation}
and
\[\frac{\partial}{\partial t} (n\langle v_{z} \rangle)+\frac{\partial}{\partial R} (n\langle v_{R}v_{z} \rangle)+\frac{1}{R}\frac{\partial}{\partial \phi} (n\langle v_{\phi}v_{z} \rangle)+\frac{\partial}{\partial z} (n\langle v^{2}_{z} \rangle)\]
\begin{equation}\label{eqa7}
+\frac{n\langle v_{R}v_{z}\rangle}{R}-n \langle a_{z} \rangle =0.
\end{equation}
\section{CALCULATION OF AVERAGED LINE OF SIGHT VELOCITY SQUARE}\label{a2}
In this part, we calculate the averaged line of sight velocity square. We assume that the central black hole is placed at the origin of our coordinates system and $ x-y $ plane is the midplane of the BLR. In this configuration, the velocity of the cloud placed at ($R, \phi, z$) can be expressed as
\begin{equation}\label{eqb1}
\mathbf{v}=v_{R}\hat{R}+v_{\phi}\hat{\phi}+v_{z}\hat{z},
\end{equation}
where $\hat{R}$, $\hat{\phi}$ and $\hat{z}$ are
\[\hat{R}=\cos \phi \hat{x}+\sin \phi \hat{y},\]
\[\hat{\phi}=-\sin \phi \hat{x}+\cos \phi \hat{y},\]
\begin{equation}\label{eqb2}
\hat{z}=\hat{z}.
\end{equation}
On the other hand, the unit vector pointing towards the observer can be written as
\begin{equation}\label{eqb3}
\hat{n}=\sin \theta_{0}\cos \phi_{0}\hat{x}+\sin \theta_{0}\sin \phi_{0}\hat{y}+\cos \theta_{0}\hat{z},
\end{equation}
where $\theta_{0}$ is the inclination angle, and $\phi_{0}$ determines the direction of $\hat{n}$. Therefore the line of sight velocity square defined by $ v_{n}^{2}=(\mathbf{v}.\hat{n})^{2}$ becomes
\[v_{n}^{2}=v_{R}^{2}\sin^{2}\theta_{0}\cos^{2}(\phi_{0} - \phi)+v_{\phi}^{2}\sin^{2}\theta_{0}\sin^{2}(\phi_{0} - \phi)+v_{z}^{2}\cos^{2}\theta_{0}\]
\[+2v_{R}v_{\phi}\sin^{2}\theta_{0}\sin(\phi_{0} - \phi)\cos (\phi_{0} - \phi)+2v_{R}v_{z}\sin\theta_{0}\cos \theta_{0}\cos (\phi_{0} - \phi)\]
\begin{equation}\label{eqb4}
+2v_{\phi}v_{z}\sin \theta_{0}\cos \theta_{0} \sin(\phi_{0} - \phi).
\end{equation}
Like subsection (\ref{ss22}) we assume $\langle v_{R}v_{z}\rangle=\langle v_{\phi}v_{z}\rangle=0 $. Thus $\langle v_{n}^{2}\rangle $ is given by
\[\langle v_{n}^{2}\rangle=\langle v_{R}^{2}\rangle \sin^{2}\theta_{0}\cos^{2}(\phi_{0} - \phi)+\langle v_{\phi}^{2}\rangle \sin^{2}\theta_{0}\sin^{2}(\phi_{0} - \phi)\]
\begin{equation}\label{eqb5}
+\langle v_{z}^{2} \rangle \cos^{2}\theta_{0}+2\langle v_{R}v_{\phi} \rangle\sin^{2}\theta_{0}\sin(\phi_{0} - \phi)\cos (\phi_{0} - \phi).
\end{equation}
Here we define the average of $\langle v_{n}^{2}\rangle $ over the $\phi $ coordinate as $\langle v_{n}^{2}\rangle_{avr}=(1/2\pi) \int_{0}^{2\pi}\langle v_{n}^{2}\rangle d\phi $. By taking the average of equation (\ref{eqb5}) and by using $\beta = \langle v_{\phi}^{2}\rangle/\langle v_{R}^{2}\rangle$, we can finally write $\langle v_{n}^{2}\rangle_{avr}$ as
\begin{equation}\label{eqb6}
\langle v_{n}^{2} \rangle _{avr}=\frac{1+\beta}{2}\langle v_{R}^{2} \rangle \sin^{2} \theta_{0}+\langle v_{z}^{2} \rangle \cos^{2} \theta_{0}.
\end{equation}
\end{document}